\documentclass[journal, onecolumn,12pt]{IEEEtran}
\usepackage{amsmath,amssymb,amsthm,lipsum} 
\usepackage{cite}
\usepackage{algorithmic}
\usepackage{algorithm}
\usepackage{graphicx}
\usepackage{textcomp}
\usepackage{stfloats}
\usepackage{booktabs}
\usepackage{bm}
\usepackage{subfigure}
\usepackage{setspace}
\usepackage{color}
\usepackage[T1]{fontenc}
\usepackage{cases}
\allowdisplaybreaks[4]

\ifCLASSINFOpdf
\else
\fi
\theoremstyle{Theorem}
\newtheorem{theo}{Theorem}
\newtheorem{theoremmaindescription1}[theo]{Theorem}
\newtheorem{theoremmaindescription2}[theo]{Theorem}
\newtheorem{theoremmaindescription3}[theo]{Theorem}
\newtheorem{theoremmaindescription4}[theo]{Theorem}
\newtheorem{theoremmaindescription5}[theo]{Theorem}

\theoremstyle{remark}
\newtheorem{rmk}{Remark}
\newtheorem{rmk2}[rmk]{Remark}

\theoremstyle{Definition}
\newtheorem{def1}{Definition}
\newtheorem{def2}[def1]{Definition}
\newtheorem{def3}[def1]{Definition}
\newtheorem{def4}[def1]{Definition}

\theoremstyle{Lemma}
\newtheorem{lemma1}{Lemma}
\newtheorem{lemma2}[lemma1]{Lemma}
\newtheorem{lemma3}[lemma1]{Lemma}
\newtheorem{lemma4}[lemma1]{Lemma}
\newtheorem{lemma5}[lemma1]{Lemma}
\newtheorem{lemma6}[lemma1]{Lemma}

\newtheorem{lemma8}[lemma1]{Lemma}

\theoremstyle{Corollary}
\newtheorem{Corollary1}{Corollary}
\newtheorem{Corollary2}[Corollary1]{Corollary}
\newtheorem{Corollary3}[Corollary1]{Corollary}
\newtheorem{Corollary4}[Corollary1]{Corollary}
\newtheorem{Corollary5}[Corollary1]{Corollary}
\newtheorem{Corollary6}[Corollary1]{Corollary}

\theoremstyle{Proposition}
\newtheorem{prop}{Proposition}
\newtheorem{proposition1}[prop]{Proposition}
\newtheorem{proposition2}[prop]{Proposition}

\begin{document}
%

%
%
%

\title{Joint Block-Sparse Recovery Using Simultaneous BOMP/BOLS}

\author{Liyang~Lu,~\IEEEmembership{Graduate Student Member,~IEEE,}
	      Zhaocheng~Wang,~\IEEEmembership{Fellow,~IEEE,} \\  
	      Sheng~Chen,~\IEEEmembership{Fellow,~IEEE} \\     
\thanks{\textit{Corresponding author}: Zhaocheng Wang.} %
\thanks{L.~Lu and Z.~Wang are with the Beijing National Research Center for Information Science and Technology, Department of Electronic Engineering, Tsinghua University, Beijing 100084, China (e-mails: luliyang@mail.tsinghua.edu.cn, zcwang@tsinghua.edu.cn), Z.~Wang is also with the Tsinghua Shenzhen International Graduate School, Shenzhen 518055, China.} %
\thanks{S.~Chen is with the School of Electronics and Computer Science, University of Southampton, Southampton SO17 1BJ, U.K. (e-mail: sqc@ecs.soton.ac.uk).} %
\vspace*{-5mm}
}

%
%

\markboth{}%
{Shell \MakeLowercase{\textit{et al.}}: Bare Demo of IEEEtran.cls for IEEE Journals}
%



\maketitle

\begin{abstract}
We consider the greedy algorithms for the joint recovery of high-dimensional sparse signals based on the block multiple measurement vector (BMMV) model in compressed sensing (CS). To this end, we first put forth two versions of simultaneous block orthogonal least squares (S-BOLS) as the baseline for the OLS framework. Their cornerstone is to sequentially check and select the support block to minimize the residual power. Then, parallel performance analysis for the existing simultaneous block orthogonal matching pursuit (S-BOMP) and the two proposed S-BOLS algorithms is developed. It indicates that under the conditions based on the mutual incoherence property (MIP) and the decaying magnitude structure of the nonzero blocks of the signal, the algorithms select all the significant blocks before possibly choosing incorrect ones. In addition, we further consider the problem of sufficient data volume for reliable recovery, and provide its MIP-based bounds in closed-form. These results together highlight the key role of the block characteristic in addressing the weak-sparse issue, i.e., the scenario where the overall sparsity is too large. The derived theoretical results are also  universally valid for conventional block-greedy algorithms and non-block algorithms by setting the number of measurement vectors and the block length to 1, respectively.
\end{abstract}


\IEEEpeerreviewmaketitle

\newpage

\section{Introduction}\label{introduction}

\IEEEPARstart{M}{ultuiple} measurement vector (MMV) problem \cite{64,68,69}, also known as joint sparse recovery in compressed sensing (CS) \cite{86,87}, aims to jointly reconstruct the sparse signal matrix $\mathbf{X}\in \mathbb{R}^{N\times E}$ from the matrix $\mathbf{Y}\in \mathbb{R}^{M\times E}$ that contains multiple measurement vectors:
\begin{equation}\label{CSmodel} 
	\mathbf{Y}=\mathbf{D}\mathbf{X}+\mathbf{N},
\end{equation}
based on the measurement matrix $\mathbf{D}\in \mathbb{R}^{M\times N}$ with $M\ll N$, where $\mathbf{N}\in \mathbb{R}^{M\times E}$ is the additive noise. It exhibits superior recovery performance over the problem based on single measurement vector (SMV) model, due to the utilization of the correlation among sparse signals and also the confirmed potential to filter out noise interference \cite{73}. In many applications, such as cooperative spectrum sensing \cite{65}, wireless channel estimation \cite{71}, two-dimensional harmonic retrieval \cite{72}, biomedical image reconstruction \cite{70}, and so on, the signals to be recovered share a common sparsity pattern, and the intrinsic issue is thus consistent with the MMV problem~(\ref{CSmodel}). 

Over the years, numerous joint sparse recovery algorithms have been proposed \cite{74,75,76,77,78}. Generally speaking, these algorithms can be divided into two categories: those based on convex optimization \cite{80} and those using greedy iterative mechanism \cite{79,68}. Although the first category provides more accurate performance, its application is limited owing to the associated high computational complexity. By contrary, the greedy one is considered to be more promising, especially when $M$ and $N$ are large, because of its fast implementation and satisfactory performance \cite{14}.

In the greedy-based algorithms, orthogonal matching pursuit (OMP) and orthogonal least squares (OLS) act as two classic frameworks and contribute to many extended algorithms for the MMV model, such as simultaneous orthogonal matching pursuit (S-OMP) \cite{79} and simultaneous multiple orthogonal least squares (S-MOLS) \cite{68}. Both OMP and OLS gradually construct an estimated support of the sparse signals by adding one or more new entries into it per iteration, and then calculate the sparse approximation over the enlarged support. The two frameworks however differ in the way of selecting the new support entry \cite{61,67}. More specifically, OMP chooses an atom that is most strongly correlated with the residual, while OLS picks a column that minimizes the power of the new residual. It can be shown that compared with the OMP framework, the OLS framework obtains better convergence characteristic, at the expense of imposing higher computational complexity \cite{66}. 

Among the variants based on the OMP/OLS frameworks, block-structure algorithms receive much attention. They exploit the inherent structure of the sparse signals where the nonzero entries appear in clusters, and obtain better provable recovery characteristic than the schemes which treat the signals being randomly sparse \cite{4}. In the applications, e.g., muti-band signal processing \cite{57,81,82}, the block-sparsity occurs naturally. Well-known approaches, such as block OMP (BOMP) \cite{4} and block OLS (BOLS) \cite{8}, have been demonstrated to have more relaxed theoretical recovery conditions than those of the conventional OMP and OLS, and they also perform well in practical implementation. For the block MMV (BMMV) problem, the authors in \cite{65} proposed a simultaneous BOMP (S-BOMP) for cooperative spectrum sensing, which effectively overcomes the impact of multi-path and fading channels. Unfortunately, there is no OLS framework-based joint block-sparse algorithm, and the reliability of the OLS framework-based algorithm when facing unsatisfactory mutual incoherence properties (MIPs) \cite{67,8} has also been ignored in the current literature. 

For the joint block-sparse signal recovery, a fundamental issue is to analyze the recoverability of the algorithms. Restricted isometry property (RIP) \cite{83,84,60} is one of the main tools for measuring the recoverability of the greedy algorithms. It indicates that if a measurement matrix $\mathbf{D}$ satisfies the RIP with some suitable restricted isometry constants (RICs), then $\mathbf{X}$ can be recovered reliably. However, calculating the RIC of a matrix is an NP-hard problem. By contrast, MIP \cite{61,62,8} is computable, and it provides stronger condition than that of the RIP \cite{3}. In other words, meeting MIP implies that RIP holds but the converse is not true. 

In the literature, there exist some known results based on MIP. In \cite{62}, the intuitive MIP-based condition for OMP's exact recovery in noiseless scenario was first provided, and the noisy recovery conditions were derived in \cite{3} by exploiting the exact recovery condition given in \cite{62}. The work \cite{60} improved the result of \cite{62} to a more relaxed one. Then the authors of \cite{61} further relaxed the result in \cite{62} based on the decaying structure of the sparse signals, and also presented the recoverability conditions in the noisy scenario. The work \cite{61} further pointed out that the sparse signals with decaying nonzero atoms appear in many applications, e.g., image and audio processing \cite{85}. For the BOMP algorithm, the authors of \cite{4} defined the new concepts of the block-MIP, and provided sufficient condition for exact recovery accordingly, while the authors of \cite{88} developed a sharp condition for stable recovery. Recently, the research in \cite{8} derived the reliable recovery conditions for a family of the OLS algorithms based on some probability assumptions, which is demonstrated to be better than the existing results. The work \cite{65} proposed the S-BOMP algorithm and extended the MIP-based sufficient conditions to the joint sparse recovery case.

Despite of the aforementioned achievements, including in algorithm designs and the corresponding theoretical analysis, the investigation of the greedy algorithms based on the BMMV model is still at its infancy, especially the ones based on the OLS framework. There are two key impediments to the study of the joint block-sparse recovery. First, since the OLS framework is known to be computationally more expensive than the OMP framework, it necessitates designing the low-complexity OLS framework-based algorithms. Second, as the BMMV model induces large data dimension, it is assignable to address this problem in the OLS-based paradigm. Currently, there is a lack of theoretical analysis framework for the BMMV model based on MIP. Compared with the conventional SMV model, BMMV further contains two additional dimensions, i.e., the block structure and the multiple measurement vectors, thus necessitating new theoretical derivation methods that go beyond the present ones. In particular, low-complexity algorithm design based on the OLS framework is waiting to be resolved, and in-depth performance analysis of the greedy joint block-sparse recovery algorithms is also required to enhance the theoretical interpretability before the algorithms can be implemented in practice. 

Against the above backgrounds, this paper sheds lights on the characterization of the recoverability for the joint block-sparse recovery algorithms in CS. The contributions are summarized as follows.

\begin{enumerate}
\item As the OLS framework exhibits more promising recovery ability to different MIP conditions than that of the OMP framework, two joint block-sparse reconstruction algorithms, called simultaneous BOLS (S-BOLS) and fast S-BOLS (FS-BOLS), are proposed. We prove that the block selection rules of the two proposed algorithms are equivalent in low coherence scenarios. The fast version thus compensates for the high computational complexity of S-BOLS for practical use. The proposed algorithms serve as the building blocks for the subsequent theoretical analysis of the joint block-sparse recovery.
		
\item The parallel performance guarantees of the S-BOMP/S-BOLS/FS-BOLS algorithms are derived. This is the first to exploit the merits of the block MMV and the decaying signal model simultaneously. The block MMV yields ample nonzero support information and block-structure property, while the decaying model enables the sparse signal matrix gradientization for stronger support power. The developed sufficient conditions indicate that if the joint sparse signal matrix obey a derived decaying gradient, the algorithms can perform reliable recovery. 
	
\item Our derived theoretical results highlight the key role of the block-structure characteristic in maintaining recovery reliability. They  reveal that the increase of block length contributes more than the increase of the number of measurements in some asymptotic cases. The results further indicate that a larger block length enables stronger capability in addressing the issues of weak-sparse recovery. Correspondingly, we develop an improved upper bound of reconstructible sparsity, and point out that this bound is a quadratic function of the block length. It is worth noting that our derived bounds based on the decaying signal structure are better than the existing ones. 
	
\item We further analyze the data volume bounds of closed-form based on the MIP of the measurement matrix. The results indicate that the number of the measurement vectors should be constrained by an upper bound, which contradicts to intuition, as more measurement vectors would appear to induce more useful information for reliable recovery. We then clarify that more measurement vectors cause stronger noise interference. The data volume is also lower bounded from the beneficial information aspect. In the noiseless scenario, this lower bound is equal to 1, which is consistent with the intuition that at least one measurement vector is required for recovery, and indicates that one measurement vector is sufficient for reliability.
\end{enumerate}

The rest of the paper is organized as follows. Section~\ref{S2} introduces the notations and BMMV background, followed by some useful definitions. In Section~\ref{S3}, we propose the S-BOLS and FS-BOLS algorithms. In Section~\ref{S4}, the performance guarantees for the S-BOMP/S-BOLS/FS-BOLS are derived. In Section~\ref{datavolume}, the analyses of the data volume are conducted, and the conclusions are presented in Section~\ref{S6}.

\section{Preliminaries}\label{S2}

\subsection{Notations}\label{S2.1}

We briefly summarize the notations used in this paper. Vectors are denoted by boldface lowercase letters, e.g., $\mathbf{x}$, and matrices are denoted by boldface uppercase letters, e.g., $\mathbf{D}$. The $i$-th element of vector $\mathbf{x}$ is denoted as $\mathbf{x}_i$. The element in the $i$-th row and $j$-th column of matrix $\mathbf{D}$ is represented by $\mathbf{D}_{ij}$, and $\mathbf{D}_{i}$ denotes the $i$-th column of $\mathbf{D}$. $\mathbf{D}_{\mathbf{\Theta}}$ is a sub-matrix of $\mathbf{D}$ that contains the columns indexed by set $\mathbf{\Theta}$. $\mathbf{D}\backslash \mathbf{D}_{\mathbf{\Theta}}$ is the residual matrix after all the columns indexed by $\mathbf{\Theta}$ are removed from $\mathbf{D}$. $\mathbf{D}^{\rm T}$ represents the transpose of $\mathbf{D}$, and $\mathbf{W}^{\rm H}$ denotes the conjugate transpose of $\mathbf{W}$. Given a block length, the $i$-th column-block submatrix of matrix $\mathbf{D}$ is denoted as $\mathbf{D}_{[i]}$, and the $i$-th row-block submatrix of $\mathbf{D}$ is denoted as $\mathbf{D}_{<i>}$. $|\mathbf{\Theta}|$ stands for the cardinality of set $\mathbf{\Theta}$ and $|c|$ is the absolute value of constant $c$. The set $\mathbf{\Theta}_B$ consists of the block indices of the set $\mathbf{\Theta}$ based on a given block length. $<\cdot,\cdot>$ denotes the inner product operation. $\mathbf{I}$ stands for the identity matrix. If $\mathbf{D}_{\mathbf{\Theta}}$ has the full column rank, $\mathbf{D}_{\mathbf{\Theta}}^\dag=\big(\mathbf{D}_{\mathbf{\Theta}}^{\rm T}\mathbf{D}_{\mathbf{\Theta}}\big)^{-1}\mathbf{D}^{\rm T}_{\mathbf{\Theta}}$ represents the pseudoinverse of $\mathbf{D}_{\mathbf{\Theta}}$. $\text{span}(\mathbf{D}_{\mathbf{\Theta}})$ denotes the space spanned by the columns of $\mathbf{D}_{\mathbf{\Theta}}$, and $\bm{P}_{\mathbf{D}_{\mathbf{\Theta}}}=\mathbf{D}_{\mathbf{\Theta}}\mathbf{D}_{\mathbf{\Theta}}^\dag$ is the projection onto $\text{span}(\mathbf{D}_{\mathbf{\Theta}})$, while $\bm{P}_{\mathbf{D}_{\mathbf{\Theta}}}^\bot=\mathbf{I}-\bm{P}_{\mathbf{D}_{\mathbf{\Theta}}}$ is the projection onto the orthogonal complement of span$(\mathbf{D}_{\mathbf{\Theta}})$. For a vector $\mathbf{r}$, ${\rm diag}(\mathbf{r})$ denotes the diagonal matrix whose diagonal elements are the entries of $\mathbf{r}$.

\subsection{BMMV Model}\label{S2.2}

The BMMV model is an extension of the MMV model (\ref{CSmodel}) where the nonzero elements of $\mathbf{X}$ occur in blocks \cite{65}. Let $d$ be the block length. Then the multiple block-sparse signal matrix is given by $\mathbf{X}=[\mathbf{X}_1,\mathbf{X}_2,\cdots,\mathbf{X}_E]\in \mathbb{R}^{N\times E}$, where $\forall i\in\{1,2,\cdots,E\}$, the $i$-th column of $\mathbf{X}$ can be expressed as
\begin{equation}\label{sparsex} 
  \mathbf{X}_i\! =\! [\underbrace{\mathbf{X}_{i1}\cdots  \mathbf{X}_{id} }_{\mathbf{X}^{\rm T}_{i[1]}} \underbrace{\mathbf{X}_{i(d+1)}\cdots \mathbf{X}_{i(2d)}}_{\mathbf{X}^{\rm T}_{i[2]}}\cdots \underbrace{\mathbf{X}_{i(N-d+1)}\cdots \mathbf{X}_{iN}}_{\mathbf{X}^{\rm T}_{i[N_B]}}]^{\rm T}\!\! ,\!
\end{equation}
in which $N\! =\! N_Bd$ and $\mathbf{X}_{i[j]}\! \in\! \mathbb{R}^{d\times 1}$ is the $j$-th block or subvector of $\mathbf{X}_i$ for $j\! \in\!\{1,2,\cdots,N_B\}$. Denote $\mathbf{X}_{<j>}\! =\! \big[\mathbf{X}_{1[j]},\mathbf{X}_{2[j]},\cdots,\mathbf{X}_{E[j]}\big]\! \in\! \mathbb{R}^{d\times E}$ as the $j$-th row-block submatrix of $\mathbf{X}$ for $j\! \in\! \{1,2,\cdots, N_B\}$, and ${\rm supp}(\mathbf{X})$ returns the set containing the indices of the nonzero row-block submatrices of $\mathbf{X}$. If there exist $k$ row-block submatrices with nonzero F-norm, $\mathbf{X}$ is called joint $k$ block-sparse. For a joint $k$ block-sparse matrix, the indices of nonzero F-norm row-block submatrices constitute the support set $\mathbf{\Omega}_B^{\star}$ with $|\mathbf{\Omega}_B^{\star}|\! =\! k$. Without loss of generality, assume that $\mathbf{\Omega}_B^{\star}\! =\! \{1,2,\cdots,k\}$ and correspondingly $\mathbf{\Omega}^{\star}\! =\! \{1,2,\cdots,kd\}$, and $\forall j\! \in\! \{1,2,\cdots,k\}$,
\begin{equation}\label{decayingarrange} 
	\|\mathbf{X}_{<1>}\|_F\geq\|\mathbf{X}_{<2>}\|_F\geq\cdots\geq\|\mathbf{X}_{<k>}\|_F>0.
\end{equation} 
Denote $\bar{\mathbf{\Omega}}_B^{\star}=\{1,2,\cdots,N_B\}\backslash\mathbf{\Omega}_B^{\star}$ as the complementary subset. Also the measurement matrix $\mathbf{D}\in\mathbb{R}^{M\times N}$ can be rewritten as a concatenation of the $N_B$ column blocks, i.e.,
\begin{equation}\label{matrixblock} 
	\mathbf{D}=[\underbrace{\mathbf{D}_1\cdots  \mathbf{D}_d }_{\mathbf{D}_{[1]}} \underbrace{\mathbf{D}_{d+1}\cdots \mathbf{D}_{2d}}_{\mathbf{D}_{[2]}}\cdots \underbrace{\mathbf{D}_{N-d+1}\cdots \mathbf{D}_{N}}_{\mathbf{D}_{[N_B]}}],
\end{equation}
where $\mathbf{D}_{[i]}\in \mathbb{R}^{M\times d}$ is the $i$-th column-block submatrix of $\mathbf{D}$. Throughout the paper, the columns in the measurement matrix are normalized to have the unit $\ell_2$-norm. 

\subsection{Useful Definitions}\label{S2.3}

RIP and MIP are the two effective frameworks for analyzing the recovery performance of the CS algorithms.  
In the following, we present the definitions of the RIP framework containing the conventional RIP and the projected RIP (P-RIP) \cite{60}, and the MIP framework consisting of the conventional matrix coherence and block-structure coherence. 

The RIP framework is introduced first.

\begin{def2}\label{rip} 
(RIP) For any $K$ sparse signal $\mathbf{x}$, the measurement matrix $\mathbf{D}$ satisfies the RIP with order $K$, if the following inequality holds:
\begin{equation}\label{conventionalRIP} 
	(1-\delta_{K})\|\mathbf{x}\|_2^2\leq \|\mathbf{D}\mathbf{x}\|_2^2\leq (1+\delta_{K})\|\mathbf{x}\|_2^2,
\end{equation}
where $\delta_K\in[0, ~ 1)$. The RIC is the minimum of $\delta_K$ that satisfies (\ref{conventionalRIP}).
\end{def2}

\begin{def1}\label{p-rip} 
(Projected RIP \cite{60}) For any signal $\mathbf{x}$, the measurement matrix $\mathbf{D}$ satisfies the P-RIP ($\underline{\delta}_{h d_1,g d_2}$, $\bar{\delta}_{h d_1,g d_2}$) if and only if $\forall \mathbf{\Psi}$, $\mathbf{\Theta}$, $\mathbf{x}_\mathbf{\Theta}$, the following inequality holds:
\begin{equation}\label{P-RIP} 
  (1-\underline{\delta}_{hd_1,gd_2})\|\mathbf{x}_\mathbf{\Theta}\|^2\leq \|\tilde{\mathbf{D}}^{\mathbf{\Psi}}_{\mathbf{\Theta}}\mathbf{x}_\mathbf{\Theta}\|^2\leq (1+\bar{\delta}_{hd_1,gd_2})\|\mathbf{x}_\mathbf{\Theta}\|^2,
\end{equation}
where $\underline{\delta}_{hd_1,gd_2}\in[0, ~ 1)$, $\bar{\delta}_{h d_1,g d_2}\in(-1,~ 1)$, $|\mathbf{\Theta}|=h d_1$, $|\mathbf{\Psi}|=g d_2$, $\mathbf{\Psi}\cap\mathbf{\Theta}=\mathbf{\emptyset}$ and $\tilde{\mathbf{D}}^{\mathbf{\Psi}}_{\mathbf{\Theta}}=\mathbf{P}^{\bot}_{\mathbf{D}_{\mathbf{\Psi}}}\mathbf{D}_{\mathbf{\Theta}}$.
\end{def1}

Note that $\bar{\delta}_{h d_1,g d_2}$ may be negative, since $\forall i\in\mathbf{\Theta}$, $\|\mathbf{P}^{\bot}_{\mathbf{D}_{\mathbf{\Psi}}}\mathbf{D}_{i}\|_2\leq1$. Also $d_1$ and $d_2$ are used to distinguish whether the block length is from $\mathbf{\Theta}$ or $\mathbf{\Psi}$, and $d_1=d_2$ actually. In the sequel, $d$ is used to represent $d_1$ and $d_2$. 

Next we present the definitions of the MIP framework.

\begin{def3}\label{definitionofcoherence} 
(Matrix coherence \cite{62}) The coherence of a matrix $\mathbf{D}$, which represents the similarity of its elements, is defined as
\begin{equation}\label{coherence} 
	\mu= \max\limits_{\forall i,j\neq i} |<\mathbf{D}_i,\mathbf{D}_j>| .
\end{equation}
\end{def3}

\begin{def4}\label{definitionofsubcoherence} 
(Block-structure coherence \cite{4}) It contains two concepts, i.e., block-coherence and sub-coherence. Let $d$ be the block length. The block-coherence of $\mathbf{D}$ is defined as
\begin{equation}\label{block coherence} 
	\mu_B = \max\limits_{\forall i,j\neq i} \frac{\|\mathbf{M}_{[i,j]}\|_2}{d},
\end{equation}
where $\mathbf{M}_{[i,j]}=\mathbf{D}^{\rm T}_{[i]}\mathbf{D}_{[j]}$, and $\mathbf{D}_{[i]}$ denotes the $i$-th column-block submatrix of $\mathbf{D}$ with the block length $d$. The sub-coherence of $\mathbf{D}$ is defined as
\begin{equation}\label{sub block coherence} 
  \nu = \max\limits_{\forall l}\max\limits_{\forall i,j\neq i} \big| \big<\mathbf{D}_{{[l]}_i},\mathbf{D}_{{[l]}_j}\big> \big|,
\end{equation}
where $\mathbf{D}_{{[l]}_i}$ is the $i$-th column of $\mathbf{D}_{[l]}$.
\end{def4}

Throughout the paper, unless otherwise stated, $d$, $\mu$, $\nu$ and $\mu_B$ represent the block length, the matrix coherence, the sub-coherence and the block-coherence, respectively.

\section{Joint Block-Sparse Recovery Algorithms}\label{S3}

The most representative joint block recovery algorithm in the greedy case is the existing S-BOMP algorithm based on the OMP framework. In S-BOMP, the algorithm chooses a block that is most strongly correlated with the signal residual matrix \cite{65}, and adds the selected index to the list per iteration. Then, it estimates the block-sparse matrix over the enlarged support matrix. In this section, we propose the S-BOLS algorithm according to the OLS framework, and further develop a fast version of the S-BOLS to compensate for the high complexity of S-BOLS. Asymptotic analysis is then conducted to demonstrate the equivalence of the block selection mechanisms in both the S-BOLS and FS-BOLS algorithms.
  
\begin{algorithm}  
\renewcommand{\algorithmicrequire}{\textbf{Input:}}
\renewcommand{\algorithmicensure}{\textbf{Output:}}
\caption{S-BOLS}
\label{alg:s-bols} 
\begin{algorithmic}[1]
	\REQUIRE $\mathbf{D}$, $\mathbf{Y}$, block sparsity level $k$, block length $d$ and residual tolerant $\zeta$
	\ENSURE $\mathbf{X}$, $\mathbf{\Theta}_B\subseteq \{1,2,\cdots,N_B\}$
	\STATE $\mathbf{Initialization}$: $l=0$, $\mathbf{R}^0=\mathbf{Y}$, $\mathbf{\Theta}^0=\emptyset$, $\mathbf{X}^0=\mathbf{0}$
	\WHILE {$l< k$ and $\|\mathbf{R}^l\|_2>\zeta$}
		\STATE Set $i^{l+1}=\arg \min\limits_{j\in\{1,\cdots,N_B\}\backslash \mathbf{\Theta}_B^{l}} \|\mathbf{P}^\bot_{\mathbf{D}_{\mathbf{\Theta}^{l}\cup \{(j-1)d+1,\cdots,jd\}}}\mathbf{Y}\|_F^2$
		\STATE Augment $\mathbf{\Theta}_B^{l+1}=\mathbf{\Theta}_B^{l}\cup{\{i^{l+1}\}}$
		\STATE Estimate $\mathbf{X}^{l+1}=\arg \min\limits_{\mathbf{X}:\;{\rm supp}(\mathbf{X})=\mathbf{\Theta}_B^{l+1}}\|\mathbf{Y}-\mathbf{D}\mathbf{X}\|_F^2$
		\STATE Update $\mathbf{R}^{l+1}=\mathbf{Y}-\mathbf{D}\mathbf{X}^{l+1}$
		\STATE $l=l+1$
	\ENDWHILE
	\STATE \textbf{return} $\mathbf{\Theta}_B=\mathbf{\Theta}_B^{l}$ and $\mathbf{X}=\mathbf{X}^{l}$
\end{algorithmic}
\end{algorithm}

\subsection{S-BOLS and FS-BOLS}\label{S-BOMP/S-BOLS} 

The proposed S-BOLS algorithm is summarized in Algorithm~\ref{alg:s-bols}. Different from S-BOMP, S-BOLS is based on the OLS framework. It has been demonstrated that the OLS framework has better convergence property \cite{66}, and is insensitive to the changes of the matrix coherence \cite{67}. As presented in the third step of Algorithm~\ref{alg:s-bols}, the block matrix decreasing the residual most is selected per iteration. Note that BOLS is an extended algorithm for the SMV model based on the OLS framework, and a variant of BOLS's block selection step that facilitates the analysis of the performance guarantees is provided in \cite[Proposition 1]{8}. The support selection variant of our S-BOLS is given directly in {\textbf{Proposition~\ref{proposition1}} below. The derivation of {\textbf{Proposition~\ref{proposition1}} is similar to that of \cite[Proposition 1]{8}.

\begin{proposition1}\label{proposition1}
Define 
\begin{align}\label{ctri} 
  F^{l} \triangleq \arg\max\limits_{j_B\in\{1,..,N_B\}\backslash\mathbf{\Theta}_B^{l}} \Bigg(\frac{\|\mathbf{D}^{\rm T}_{(j_B-1)d+1} \mathbf{R}^{l}\|_F}{\|\mathbf{P}^\bot_{\mathbf{D}_{\mathbf{\Theta}^{l}}} \mathbf{D}_{(j_B-1)d+1}\|_2}\Bigg)^2 
	 +\!\!\!\! \sum^{j_B d}_{j=(j_B-1)d+2}\!\! \Bigg(\frac{\|\mathbf{D}^{\rm T}_j \mathbf{R}^{l}\|_F}{\|\mathbf{P}^\bot_{\mathbf{D}_{\mathbf{\Theta}^{l}\cup\{(j_B-1)d+1,\cdots,j-1\}}} \mathbf{D}_j\|_2}\Bigg)^2\!\! .\!
\end{align}
In the $(l+1)$-th iteration, the S-BOLS algorithm identifies the index of a column-block submatrix of the measurement matrix using the selection mechanism: $i^{l+1}=F^{l}$.
\end{proposition1}
 
The calculation of $d$ different projection matrices in (\ref{ctri}) increases the computational complexity, and the third step in Algorithm~\ref{alg:s-bols} also involves the calculation of multiple high-dimensional
projection matrices, which is the root cause for the S-BOLS algorithm to have high complexity. To ease the complexity burden for practical implementation, we modify the S-BOLS algorithm into a fast version called FS-BOLS which is presented in Algorithm~\ref{alg:Fs-bols}. More specifically, define
\begin{align}\label{new-ctri} 
	F^{l}_r\triangleq \arg \! \max\limits_{j_B\in\{1,..,N_B\}\backslash\mathbf{\Theta}_B^{l}}\! \sum^{j_B d}_{j=(j_B-1)d+1}\!\!\! \Bigg(\frac{\|\mathbf{D}^{\rm T}_j \mathbf{R}^{l}\|_F}{\|\mathbf{P}^\bot_{\mathbf{D}_{\mathbf{\Theta}^{l}}}\mathbf{D}_j\|_2}\Bigg)^2\!\! .\!
\end{align}
Then the FS-BOLS algorithm selects a new support using $i^{l+1}=F^{l}_r$, rather than using $i^{l+1}=F^{l}$ as in the S-BOLS. 

\begin{algorithm}
\renewcommand{\algorithmicrequire}{\textbf{Input:}}
\renewcommand{\algorithmicensure}{\textbf{Output:}}
\caption{FS-BOLS}
\label{alg:Fs-bols} 
\begin{algorithmic}[1]
	\REQUIRE $\mathbf{D}$, $\mathbf{Y}$, block sparsity level $k$, block length $d$ and residual tolerant $\zeta$
	\ENSURE $\mathbf{X}$, $\mathbf{\Theta}_B\subseteq \{1,2,\cdots,N_B\}$
	\STATE $\mathbf{Initialization}$: $l=0$, $\mathbf{R}^0=\mathbf{Y}$, $\mathbf{\Theta}^0=\emptyset$, $\mathbf{X}^0=\mathbf{0}$
	\WHILE {$l< k$ and $\|\mathbf{R}^l\|_2>\zeta$}
		\STATE Calculate $F_r^l$ of (\ref{new-ctri}) and set $i^{l+1}=F_r^l$
		\STATE Augment $\mathbf{\Theta}_B^{l+1}=\mathbf{\Theta}_B^{l}\cup{\{i^{l+1}\}}$
		\STATE Estimate $\mathbf{X}^{l+1}=\arg \min\limits_{\mathbf{X}:\;{\rm supp}(\mathbf{X})=\mathbf{\Theta}_B^{l+1}}\|\mathbf{Y}-\mathbf{D}\mathbf{X}\|_F^2$
		\STATE Update $\mathbf{R}^{l+1}=\mathbf{Y}-\mathbf{D}\mathbf{X}^{l+1}$
		\STATE $l=l+1$
	\ENDWHILE
	\STATE \textbf{return} $\mathbf{\Theta}_B=\mathbf{\Theta}_B^{l}$ and $\mathbf{X}=\mathbf{X}^{l}$
\end{algorithmic}
\end{algorithm}

Intuitively, (\ref{new-ctri}) is different from (\ref{ctri}) but it imposes a lower complexity than the latter. In what follows, the condition for (\ref{new-ctri}) and (\ref{ctri}) to be equivalent is presented. First we present a lemma of the eigenvalue bounds of a given matrix, which is an extension of \cite[Lemma 2]{3} in a block-structure manner.

\begin{lemma1}\label{lemma1}
Given a matrix $\mathbf{D}\in\mathbb{R}^{M\times k d}$ that consists of $k$ column-block submatrices of size ${M\times d}$, define $\lambda_{\min}$ and $\lambda_{\max}$ as the minimum and maximum eigenvalues of the matrix $\mathbf{D}^{\rm T}\mathbf{D}\in\mathbb{R}^{kd\times kd}$. When $(d-1)\nu+(k-1)d\mu_B<1$, the following inequality holds:
\begin{equation}\label{eigenvalue-bounds} 
	\begin{aligned}
		1-(d-1)\nu  -(k-1)d\mu_B\leq\lambda_{\min} 
		 \leq\lambda_{\max}\leq1+(d-1)\nu+(k-1)d\mu_B,
	\end{aligned}    
\end{equation}
where $\nu$ and $\mu_B$ are the sub-coherence and block-coherence of $\mathbf{D}$, respectively.
\end{lemma1}

\begin{IEEEproof}
See Appendix~\ref{proofoflemma1}.
\end{IEEEproof}

\begin{proposition2}\label{Props2}
Asymptotically, if the MIP of the Gaussian measurement matrix meets its lower bound, $F^{l}$ and $F^{l}_r$ satisfy:
\begin{align}\label{eqProps2} 
	\lim\limits_{M/N=\omega;\;M,N\rightarrow\infty}F^{l}\equiv F^{l}_r ,
\end{align}
where $\omega$ is a fixed parameter known as the compression rate.
\end{proposition2}

\begin{IEEEproof}
See Appendix~\ref{proofofprops2}.
\end{IEEEproof}

Another reachable lower bound of $\mu_B$ with respect to the block length $d$ and the number of measurements $M$ is \cite{4} 
\begin{equation}\label{miublower1} 
	\mu_B\geq \frac{1}{\sqrt{dM}}.
\end{equation}
An example that achieves the lower bound in the inequality (\ref{miublower1}) \cite{4} is presented here. Let $\mathbf{H}$ denote the DFT matrix of size $m\times m$ with $\mathbf{H}_{ij}=(1/\sqrt{m})e^{\textsf{j}2\pi ij/m}$. Define $\mathbf{W}=\mathbf{I}$ and $\mathbf{Q}=\mathbf{H}\otimes\mathbf{U}$, where $\otimes$ denotes the Kronecker product and $\mathbf{U}$ is a unitary matrix. Then, $\mathbf{W}^{\rm H}_{[i]}\mathbf{Q}_{[j]}=\mathbf{H}_{ij}\mathbf{U}$, and $\mu_B=1/\sqrt{dM}$ \cite{4}. We call $\mathbf{W}$ and $\mathbf{Q}$ the block spike-Fourier pair. Suppose that $d/M$ and $M/N$ are fixed to $\omega_1$ and $\omega_2$, respectively. Then the block-coherence of the block spike-Fourier pair satisfies
\begin{align} 
	 \lim_{d/M=\omega_1;M/N=\omega_2;M,N\rightarrow\infty} \mu_B  
	 = \lim_{d/M=\omega_1;M/N=\omega_2;M,N\rightarrow\infty} \frac{1}{\sqrt{d M}} = 0 .\label{miublower2}
\end{align} 
Moreover, based on $\mu_B=1/(\sqrt{d M})$, we also have
\begin{equation}\label{miublower4} 
 	\lim_{d\rightarrow\infty} \mu_B = \lim_{d\rightarrow\infty} \frac{1}{\sqrt{d M}} = 0 .
\end{equation}
A larger $d$ induces stronger block-structure characteristics, leading to better block MIP. When $d=1$, the pair of $\mathbf{W}$ and $\mathbf{Q}$ reduce to the conventional spike-Fourier pair, which is proved to yield the most mutually incoherence \cite{63} and $\mu=1/(\sqrt{M})$. Similar to (\ref{miublower2}), we have
\begin{equation}\label{miublower3} 
	\lim_{M/N=\omega_2;M,N\rightarrow\infty} \mu= \lim_{M/N=\omega_2;M,N\rightarrow\infty} \frac{1}{\sqrt{M}} = 0 .
\end{equation} 
These practical paradigms prove the actual rationality of \textbf{Proposition~\ref{Props2}} again. 

\textbf{Proposition~\ref{Props2}} demonstrates the equivalence between $F^{l}$ and $F^{l}_r$ in the asymptotic case, the following analysis presents the computational complexity advantage of $F^{l}_r$ over that of $F^{l}$.

\begin{rmk}\label{Remark1}
The main difference between $F^{l}$ and $F^{l}_r$ lies in the calculation of the orthogonal projection matrix. Note that the computational complexity of computing $\mathbf{P}^\bot_{\mathbf{D}_{\mathbf{\Theta}^{l}}}$ is on the order of $\mathcal{O}\big(|\mathbf{\Theta}^{l}|^3+|\mathbf{\Theta}^{l}|^2M+M^2\big)$. Suppose that both the S-BOLS and FS-BOLS iterate $k$ times. Then, the total computational complexity of computing the projection matrices in the algorithms using $F^{l}$ and $F^{l}_r$ are  
\begin{align} 
  L(F^{l}) =& \mathcal{O}\bigg(\sum_{l=1}^{k-1}\sum_{i=0}^{d-1}|l+i|^3+|l+i|^2M+M^2\bigg) , \label{complexity of F} \\
	L(F^{l}_r) =& \mathcal{O}\bigg(\sum_{l=1}^{k-1}|l|^3+|l|^2M+M^2\bigg), \label{complexity of Fr}
\end{align}
respectively. Asymptotically, we have
\begin{align} 
	\lim\limits_{M\rightarrow\infty}	L(F^{l}) =& \mathcal{O}\big((k-1)dM^2\big), \label{complexity of Fm} \\
	\lim\limits_{M\rightarrow\infty}L(F^{l}_r) =& \mathcal{O}\big((k-1)M^2\big). \label{complexity of Frm}
\end{align}
Intuitively, $F^{l}_r$ has a $d$-times complexity advantage. In block-sparse recovery research, such as \cite{4,8}, it declares that when the other conditions are fixed, the larger the block length, the better the performance. Since (\ref{complexity of Frm}) is independent of the block length $d$, the FS-BOLS can provide a reliable recovery while maintaining the advantage of low complexity when the block length increases. 
\end{rmk}

\subsection{Simple Block Selection Mechanism Expressions}\label{S3.2}

Based on the descriptions in Subsection~\ref{S-BOMP/S-BOLS}, we present the expressions of the support selection rules for the S-BOMP/S-BOLS/FS-BOLS, which will be used to facilitate the theoretical analysis in the following sections.

For the $(l+1)$-th iteration, the support selection rule of the S-BOMP is \cite{65}
\begin{equation}\label{omp choose rule} 
 	j\in \arg \max_{i\notin \mathbf{\Omega}_B^l} \big\|\mathbf{D}_{[i]}^{\rm T}\mathbf{R}^{l}\big\|_F,
\end{equation}
where the set $\mathbf{\Omega}_B^l$ contains the block indices of the selected supports during the last $l$ iterations. 

The S-BOLS seeks a support that provides the most significant decrease in the residual power, as shown in \textbf{Proposition~\ref{proposition1}}, while the FS-BOLS is asymptotically equivalent to the S-BOLS as stated in \textbf{Proposition~\ref{Props2}}. The selection rules of the S-BOLS/FS-BOLS can be expressed in terms of the projected submatrices of the measurement matrix \cite{61}. Without loss of generality, let the indices selected during the $(l+1)$-th iteration be $(1,\cdots,d)$. For the $(l+1)$-th iteration, denote
\begin{align} 
	\tilde{\mathbf{D}}_{[i]} \triangleq & \mathbf{P}_{\mathbf{D}_{\mathbf{\Omega}^l}}^{\bot}\mathbf{D}_{[i]}, \label{simplerule1} \\
	\tilde{\mathbf{B}}_{[i]} \triangleq & \left\{\begin{array}{cl}
		\tilde{\mathbf{D}}_{[i]}\mathbf{T},  & \enspace \text{if} \enspace \tilde{\mathbf{D}}_{[i]}\neq\mathbf{0}\enspace \text{for S-BOLS}, \\
		\tilde{\mathbf{D}}_{[i]}\mathbf{T}_f, & \enspace \text{if}\enspace \tilde{\mathbf{D}}_{[i]}\neq\mathbf{0}\enspace \text{for FS-BOLS}, \\
		\mathbf{0}, & \enspace \text{otherwise},
	\end{array} \right. \label{simplerule2}
\end{align}
where 
\begin{align} 
	&\mathbf{T} = {\rm diag}\bigg(  \frac{1}{\|\mathbf{P}^\bot_{\mathbf{D}_{\mathbf{\Omega}^l}}\mathbf{D}_1\|_2},\frac{1}{\|\mathbf{P}^\bot_{\mathbf{D}_{\mathbf{\Omega}^l\cup\{1\}}}\mathbf{D}_{2}\|_2},\cdots,  
	 \frac{1}{\|\mathbf{P}^\bot_{\mathbf{D}_{\mathbf{\Omega}^l\cup\{1,\cdots,d-1\}}}\mathbf{D}_{d}\|_2}\bigg), \label{Rodefinition} \\
	&\mathbf{T}_f ={\rm diag} \bigg(  \frac{1}{\|\mathbf{P}^\bot_{\mathbf{D}_{\mathbf{\Omega}^l}}\mathbf{D}_1\|_2},\frac{1}{\|\mathbf{P}^\bot_{\mathbf{D}_{\mathbf{\Omega}^l}}\mathbf{D}_{2}\|_2},\cdots,  
	 \frac{1}{\|\mathbf{P}^\bot_{\mathbf{D}_{\mathbf{\Omega}^l}}\mathbf{D}_{d}\|_2}\bigg). \label{Rodefinitionfs}
\end{align}

Based on the above descriptions, the selection rules of the S-BOMP/S-BOLS/FS-BOLS algorithms in the $(l+1)$-th iteration can be reformulated jointly as
\begin{align}\label{rulewhere} 
	j\in \arg \max_{i\notin \mathbf{\Omega}_B^l} \big\|\tilde{\mathbf{C}}_{[i]}^{\rm T}\mathbf{R}^l\big\|_F,
\end{align}
with	
\begin{align}\label{s-bomp choose rule} 
	\tilde{\mathbf{C}}_{[i]} \triangleq \left\{\begin{array}{cl}
		\tilde{\mathbf{D}}_{[i]}, & \enspace\text{for S-BOMP}, \\
		\tilde{\mathbf{B}}_{[i]}, & \enspace\text{for S-BOLS/FS-BOLS} .
	\end{array}\right.
\end{align}

\section{Performance Guarantees}\label{S4}

In this section, we study the recovery performance of the S-BOMP/S-BOLS/FS-BOLS algorithms under the RIP and MIP frameworks. We first provide some useful lemmas as the theoretical basis for the subsequent derivations. Then, the main recovery conditions are developed, which ensure the algorithms select correct supports.

\subsection{Some Useful Lemmas}\label{S4.1}

First, we introduce the following lemma describing the spectral norm bounds of two different matrices. \setcounter{equation}{28}

\begin{lemma3}\label{lemma3} 
Denote the block-coherence and block length of a block-structure matrix $\mathbf{D}$ as $\mu_B$ and $d$, respectively. Express two sub-matrices of $\mathbf{D}$ as $\mathbf{D}_{\mathbf{\Theta}}\in\mathbb{R}^{M\times g d}$ and $\mathbf{D}_{\mathbf{\Psi}}\in\mathbb{R}^{M\times h d}$, where $|\mathbf{\Theta}|=g d$, $|\mathbf{\Psi}|=h d$, and $\mathbf{\Theta}\cup\mathbf{\Psi}=\mathbf{\emptyset}$. Then, the following inequality holds:
\begin{equation}\label{lemma3main} 
	\big\|\mathbf{D}^{\rm T}_{\mathbf{\Theta}}\mathbf{D}_{\mathbf{\Psi}}\big\|_2\leq \min\Big\{g d^{\frac{3}{2}}\mu_B, h d^{\frac{3}{2}}\mu_B\Big\}.
\end{equation}
\end{lemma3}

\begin{IEEEproof}
See Appendix~\ref{proofoflemma3}.
\end{IEEEproof}

\begin{lemma2}\label{lemma2} 
If $(d-1)\nu+(g-1)d\mu_B < 1$, then $\mathbf{D}$ satisfies the P-RIP$(\underline{\delta}_{h d,g d}$, $\bar{\delta}_{hd,gd})$ for any $h\geq0$, $d>1$, where 
\begin{align}\label{anylowerP-RIP} 
	\underline{\delta}_{hd,gd} = (d-1)\nu + (h-1)d\mu_B  
	 +\frac{\min\big\{g^2d^3\mu_B^2, h^2d^{3}\mu_B^2\big\}}{1-(d-1)\nu - (g-1)d\mu_B} ,
\end{align}
\begin{equation}\label{anyhigherP-RIP} 
	\bar{\delta}_{hd,gd} = (d-1)\nu+(h-1)d\mu_B .
\end{equation}
\end{lemma2}

\begin{IEEEproof}
See Appendix~\ref{proofoflemma2}.
\end{IEEEproof}

\begin{lemma5}\label{lemma5} 
Suppose that $\mathbf{D}$ satisfies the P-RIP$(\underline{\delta}_{2,2g}, \bar{\delta}_{2,2g})$. Then, the following inequality holds:
\begin{equation}\label{lemma5 main} 
	\|\tilde{\mathbf{D}}^{\rm T}_{[i]}\tilde{\mathbf{D}}_{[j]}\|_F\leq \frac{d(\underline{\delta}_{2,2 g}+\bar{\delta}_{2,2 g})}{2}.
\end{equation}
\end{lemma5}

\begin{IEEEproof}
The result immediately follows from the definition of F-norm and the proof of \cite[Lemma 10]{60}.
 \end{IEEEproof}

\begin{lemma6}\label{lemma6} 
Let the set $\mathbf{\Omega}^l$ consist of the selected indices during the previous $l$ $(1\leq l<k)$ iterations, and $d>1$. For the $(l+1)$-th iteration, if $\nu+2(l-1)\mu_B<1$, we have
\begin{align} 
	& \|\tilde{\mathbf{D}}_{i}\|_2^2 \geq \frac{1-\nu-2(l-1)\mu_B-4\mu_B^2}{1-\nu-2(l-1)\mu_B}, \forall i\notin\mathbf{\Omega}^l, \label{lemma6mainmain} \\
	& |\langle\tilde{\mathbf{D}}_{i},\tilde{\mathbf{D}}_{j}\rangle| \leq \bigg(\nu+\frac{4\mu_B^2}{1-\nu-2(l-1)\mu_B}\bigg), \forall i\neq j, \label{lemma6mainmain3} \\
	& \|\tilde{\mathbf{C}}^T_{[i]}\tilde{\mathbf{D}}_{[i]}\|_F \geq \alpha_l(\mu_B,\nu,d)>0, \forall i\notin \mathbf{\Omega}_B^l, \label{lemma6main1} \\
	& \|\tilde{\mathbf{C}}^T_{[i]}\tilde{\mathbf{D}}_{[j]}\|_F \leq \beta_l(\mu_B,\nu,d), \forall i\neq j, \label{lemma6main2}
\end{align}
where 
\begin{equation}\label{lemma6main4} 
	\alpha_l(\mu_B,\nu,d) = \sqrt{\frac{d(1-\nu-2(l-1)\mu_B-4\mu_B^2)}{1-\nu-2(l-1)\mu_B}},
\end{equation}	 
\begin{equation}\label{lemma6main6} 
	\beta_l(\mu_B,\nu,d) = \left\{\begin{matrix}		
	\min\bigg\{d,	d\Big(\nu+\frac{4\mu_B^2}{1-\nu-2(l-1)\mu_B}\Big)\bigg\}, \enspace \text{for S-BOMP}, \\
	\min\bigg\{d,d\sqrt{\frac{1-\nu-2(l-1)\mu_B}{1-\nu-2(l-1)\mu_B-4\mu_B^2}}\Big(\nu+\frac{4\mu_B^2}{1-\nu-2(l-1)\mu_B}\Big)\bigg\}, \enspace \text{for S-BOLS/FS-BOLS}.
\end{matrix}\right.
\end{equation}
\end{lemma6}

\begin{IEEEproof}
See Appendix~\ref{proofoflemma6}.
\end{IEEEproof}\setcounter{equation}{38}

It is interesting to compare the results given in \textbf{Lemma~\ref{lemma6}} with some of the existing results that are based on the non-structure coherence in the literature. In \cite{61}, the authors show that if $\mu<1/(ld)$, $\|\tilde{\mathbf{D}}_{i}\|_2^2\geq \frac{(\mu+1)(1-ld\mu)}{1-(ld-1)\mu}$. Since $\mu_B\leq\mu$ and $\nu<\mu$, when $d>1$, the condition $\nu+2(l-1)\mu_B<1$ is more relaxed than the condition $\mu<1/(ld)$ thanks to the introduction of block-structure. Furthermore, the lower bound of $\|\tilde{\mathbf{D}}_{i}\|_2^2$ in \textbf{Lemma~\ref{lemma6}} is closer to 1 than the bound $\frac{(\mu+1)(1-ld\mu)}{1-(ld-1)\mu}$ derived in \cite{61}.

The following analysis provides the concise expressions of $\beta_l(\mu_B,\nu,d)$ first, and then demonstrates that $\alpha_l(\mu_B,\nu,d)>\beta_l(\mu_B,\nu,d)$ based on asymptotic analysis. It is worth noting that the inequality  $\alpha_l(\mu_B,\nu,d)>\beta_l(\mu_B,\nu,d)$ plays a key role in the derivation of the recovery conditions.

Note that $\lim\limits_{\nu\rightarrow 0}\beta_l(\mu_B,\nu,d)=\beta_l(\mu_B,0,d)$. For the S-BOMP, since $\mu_B\leq 1$, we have $\frac{4\mu_B^2d}{1-2(l-1)\mu_B}\leq d$ and thus
\begin{equation}\label{analysis11} 
 	\beta_l(\mu_B,0,d) = \frac{4\mu_B^2d}{1-2(l-1)\mu_B}.
\end{equation}
For the S-BOLS/FS-BOLS, letting  
\begin{equation}\label{analysis33} 
	\sqrt{\frac{1-2(l-1)\mu_B}{1-2(l-1)\mu_B-4\mu_B^2}}\Big(\frac{4\mu_B^2}{1-2(l-1)\mu_B}\Big)\leq 1,
\end{equation}
we get 
\begin{equation}\label{analysis4} 
	16\mu_B^4\leq \big(1-2(l-1)\mu_B-4\mu_B^2\big)\big(1-2(l-1)\mu_B\big).
\end{equation}
Since $\lim\limits_{\mu_B\rightarrow 0}16\mu_B^4=0$ and $\lim\limits_{\mu_B\rightarrow0}\big(1-2(l-1)\mu_B-4\mu_B^2\big)\big(1-2(l-1)\mu_B\big)=1$, (\ref{analysis4}) holds under small block-coherence cases. Hence, (\ref{analysis33}) follows from these analyses, which means that for the S-BOLS/FS-BOLS algorithms, 
\begin{equation}\label{betanewres} 
	\beta_l(\mu_B,0,d) =\! \sqrt{\frac{1-2(l-1)\mu_B}{1-2(l-1)\mu_B-4\mu_B^2}}\Big(\frac{4\mu_B^2d}{1-2(l-1)\mu_B}\Big).
\end{equation}
Now let 
\begin{equation}\label{analysis5} 
	\alpha_l(\mu_B,0,d) =\! \sqrt{\frac{d(1-2(l-1)\mu_B-4\mu_B^2)}{1-2(l-1)\mu_B}} > \beta_l(\mu_B,0,d),
\end{equation}
we obtain
\begin{equation}\label{analysis6} 
	16\mu_B^4d < \big(1-2(l-1)\mu_B-4\mu_B^2\big)^2.
\end{equation}
Similar to the analysis of (\ref{analysis4}), for fixed $d$, we have $\alpha_l(\mu_B,0,d)>\beta_l(\mu_B,0,d)$. 

Furthermore, in the asymptotic condition that $\nu$ approaches 0, we have
\begin{align} 
  \lim\limits_{\mu_B\rightarrow0}\alpha_l(\mu_B,\nu,d) =& \sqrt{d}, \label{otheranalysis1} \\
	\lim\limits_{\mu_B\rightarrow0}\beta_l(\mu_B,\nu,d) =& d\nu\leq d. \label{otheranalysis2}
\end{align}
Then, if $\nu<\sqrt{1/d}$, $\alpha_l(0,\nu,d)>\beta_l(0,\nu,d)$. The case that $\mu_B$ and $\nu$ approach 0 simultaneously is presented as follows:
\begin{align}  
	\lim\limits_{\mu_B,\nu\rightarrow0}\alpha_l(\mu_B,\nu,d) =& \sqrt{d}, \label{otheranalysis3} \\
	\lim\limits_{\mu_B,\nu\rightarrow0}\beta_l(\mu_B,\nu,d) =& 0. \label{otheranalysis4}
\end{align}
Obviously, $\alpha_l(0,0,d)>\beta_l(0,0,d)$.

\begin{rmk2}\label{Rmark2}
Consider the most mutually incoherent case, i.e., the spike-Fourier base pair. Since $\nu\leq\mu$ \cite{4}, we have
\begin{equation}\label{rmk2main} 
	\nu = \mu = \frac{1}{\sqrt{M}}.
\end{equation}
As stated in Subsection~\ref{S-BOMP/S-BOLS}, the block-coherence of the block spike-Fourier base pair satisfies $\mu_B=1/\sqrt{d M}$. Then, we have
\begin{align} 
	\lim\limits_{d\rightarrow\infty}\alpha_l(\mu_B,\nu,d) =& \sqrt{d}, \label{rmk2main2} \\
	\lim\limits_{d\rightarrow\infty}\beta_l(\mu_B,\nu,d) =& \frac{d}{\sqrt{M}}. \label{rmk2main3}
\end{align}
It can be concluded that if $M>d$, then $\alpha_l(\mu_B,\nu,d)>\beta_l(\mu_B,\nu,d)$. Meanwhile, the authors of \cite{14} prove that $M \geq 2 k d\ln(N/\eta)$ is sufficient to guarantee the reliable recovery, where $k$ is the block sparsity, $N$ denotes the length of the sparse signal, and $\eta$ is a small constant. In the scenario that the necessary number of measurements is satisfied for reliable recovery, i.e.,  $M\geq 2 k d\ln(N/\eta)$, since this bound is also concluded in the condition $M > d$, $\alpha_l(\mu_B,\nu,d)>\beta_l(\mu_B,\nu,d)$ is easy to establish in this case.
\end{rmk2}

The aforementioned lemmas are the building blocks of the following lemma, which gives a sufficient condition for the S-BOMP/S-BOLS/FS-BOLS algorithms to choose a correct support during a given iteration. For simplicity, $\alpha_l(\mu_B,\nu,d)$ and $\beta_l(\mu_B,\nu,d)$ are denoted by $\alpha_l$ and $\beta_l$ in the sequel.

\begin{lemma4}\label{lemma4} 
Consider the system model (\ref{CSmodel}). Suppose that $\mathbf{X}\in\mathbb{R}^{N\times E}$ is the joint block-sparse matrix with the row-block submatrices $\mathbf{X}_{<i>}\in\mathbb{R}^{d\times E}$ for $1\leq i\leq N/d=N_B$, $\mathbf{\Omega}^{\star}_B$ is the block index set with $|\mathbf{\Omega}^{\star}_B|=k$, $\|\mathbf{N}_i\|_2\leq\epsilon$, and the S-BOMP/S-BOLS/FS-BOLS algorithms have selected the supports in $\mathbf{\Omega}^l_B\subseteq\mathbf{\Omega}_B^{\star}$ during the first $l$ $(1\leq l<k)$ iterations. For the $(l+1)$-th iteration, if 
\begin{align}\label{lemma4main1} 
	\nu+2(l-1)\mu_B <& 1,
\end{align}
\begin{align}\label{lemma4main2} 
	 (\alpha_l+\beta_l)\|\mathbf{X}_{<t>}\|_F-2\beta_l\sum_{i\in\mathbf{\Omega}_{B}^{\star}\backslash\mathbf{\Omega}^l_{B}}\|\mathbf{X}_{<i>}\|_F  
	 > 2\sqrt{d E}\epsilon + (1+\beta_l) \sum_{i\in\bar{\mathbf{\Omega}}_{B}^{\star}}\|\mathbf{X}_{<i>}\|_F, 
\end{align}
where $t\in\arg \max\limits _{i\in\mathbf{\Omega}^{\star}_B\backslash\mathbf{\Omega}_B^l}\|\mathbf{X}_{<i>}\|_F$,
then the S-BOMP/S-BOLS/FS-BOLS algorithms choose a support in $\mathbf{\Omega}_B^{\star}\backslash\mathbf{\Omega}^l_B$.
\end{lemma4}

\begin{IEEEproof}
See Appendix~\ref{proofoflemma4}.
\end{IEEEproof}

\subsection{Main Recovery Conditions}\label{mainconditons} 

We first derive a sufficient condition for the establishment of \textbf{Lemma~\ref{lemma4}}. Based on this result, we develop a condition guaranteeing that the S-BOMP/S-BOLS/FS-BOLS reliably recover the joint $k$ block-sparse signals in asymptotic case.

\begin{theoremmaindescription1}\label{theoremmain1} 
Suppose that $\forall e\in\{1,2,\cdots,E\}$, $\|\mathbf{N}_e\|_2\leq\epsilon$, $\mathbf{\Omega}_B^{\star}$ is the index set with $|\mathbf{\Omega}_B^{\star}|=k$, and the S-BOMP/S-BOLS/FS-BOLS algorithms have selected supports in $\mathbf{\Omega}_B^l\subseteq\mathbf{\Omega}_B^{\star}$ during the first $l$ $(1\leq l<k)$ iterations. For the $(l+1)$-th iteration, if
\begin{align} 
	 \nu+2(l-1)\mu_B < 1 , \label{theomain1} 
	 \end{align}
 \begin{align}
	 \|\mathbf{X}_{<t>}\|_F > \frac{2(k-l-1)}{\rho-1}\|\mathbf{X}_{<t+1>}\|_F  
	 +\frac{2\sqrt{d E}\epsilon+(1+\beta_l)\sum_{i\in\bar{\mathbf{\Omega}}_B^{\star}}\|\mathbf{X}_{<i>}\|_F}{\alpha_l-\beta_l}, \label{theomain2}
\end{align}
\textbf{Lemma~\ref{lemma4}} holds, where $t$ denotes the smallest index such that
\begin{equation}\label{theoprof1} 
	t\in\arg\max\limits_{i\in\mathbf{\Omega}_B^{\mathbf{\star}}\backslash\mathbf{\Omega}_B^l} \|\mathbf{X}_{<i>}\|_{F},
\end{equation}
and
\begin{equation}\label{theomain3} 
  \rho = \frac{\sqrt{1-\nu-2(l-1)\mu_B-4\mu_B^2}\sqrt{1-\nu-2(l-1)\mu_B}}{\sqrt{d}(\nu-\nu^2-2(l-1)\mu_B\nu+4\mu_B^2)}.
\end{equation}
\end{theoremmaindescription1}

\begin{IEEEproof}
See Appendix~\ref{proofoftheorem1}.
\end{IEEEproof}

It is useful to compare our results with some of the known results in the existing literature. We first turn to the upper bound of the reconstructible sparsity, that presents the sparsity bound under which the algorithms can perform reliable recovery. The bound based on the conventional coherence given in \cite{62} is 
\begin{equation}\label{reconstructible3} 
	K < \frac{1}{2}\bigg(\frac{1}{\mu}+1\bigg),
\end{equation}
where $K=k d$ represents the total sparsity. In \cite{4}, the authors present the upper bound of the reconstructible sparsity based on the block-structure characteristics as
\begin{equation}\label{reconstructible1} 
	k d < \frac{1}{2}\bigg(\frac{1}{\mu_B}+d-\frac{d\nu}{\mu_B}+\frac{\nu}{\mu_B}\bigg).
\end{equation}
On the other hand, the condition (\ref{theomain1}) in \textbf{Theorem~\ref{theoremmain1}} provides a new bound of the reconstructible sparsity level as 
\begin{equation}\label{reconstructible2} 
 	k d < \frac{1}{2}\bigg(\frac{d}{\mu_B}+2 d-\frac{d\nu}{\mu_B}\bigg)
\end{equation}
with the precondition (\ref{theomain2}). For $d=1$, and thus $\nu=0$ due to the definition of $\nu$, the upper bound in (\ref{reconstructible1}) becomes the same as (\ref{reconstructible3}), and (\ref{reconstructible2}) becomes
\begin{equation}\label{reconstructible1cc} 
	k < \frac{1}{2}\bigg(\frac{1}{\mu}+2\bigg).
\end{equation}
It can be observed that (\ref{reconstructible1cc}) is better than (\ref{reconstructible3}). This indicates that our result allows the algorithms to achieve reliable recovery under a more unsatisfactory sparse signal, i.e., the weak-sparse signal, without block-structure. For $d\geq 2$, similar conclusions can be obtained. In this case, (\ref{reconstructible2}) unveils a larger block sparsity bound for the algorithms to provide reliable reconstruction under this sparsity condition. 

Next consider a block orthogonal matrix, i.e., $\nu=0$. Then, (\ref{reconstructible1}) and (\ref{reconstructible2}) become
\begin{align} 
 	k d < \frac{1}{2}\bigg(\frac{1}{\mu_B}+d\bigg), \label{reconstructible1change} \\
	k d < \frac{1}{2}\bigg(\frac{d}{\mu_B}+2d\bigg). \label{reconstructible2change}
\end{align}
These results reveal that our result (\ref{reconstructible2change}) provides a larger reconstructible sparsity bound for block orthogonal matrix. More specifically, the larger the block length $d$, the larger the upper bound of the reconstructible sparsity, and the bound in (\ref{reconstructible2change}) grows faster with respect to $d$ than that in (\ref{reconstructible1change}).

Denote the right-side of (\ref{reconstructible2change}) as $K^{*}(\mu_B,d)$. For a Gaussian measurement matrix of size $M\times N$, the upper bound of $K^{*}(\mu_B,d)$ with respect to $\mu_B$ is $K^{*}\big(\frac{1}{d}\sqrt{\frac{N-M}{M(N-1)}},d\big)$ since $\mu_B\geq\mu/d$ \cite{8}, and $K^{*}(\mu_B,d)$ is monotonically decreasing as $\mu_B$ increases. Then, we have
\begin{align}\label{analysis1} 
	& \lim\limits_{M/N=\omega;\;M,N\rightarrow\infty}K^{*}\bigg(\frac{1}{d}\sqrt{\frac{N-M}{M(N-1)}},d\bigg)\nonumber \\
	& =	\lim\limits_{M/N=\omega;\;M,N\rightarrow\infty}K^{*}\bigg(\frac{1}{d}\sqrt{\frac{1-\omega}{M}},d\bigg)\nonumber \\
	& = \frac{1}{2}\bigg(d^2\sqrt{\frac{M}{1-\omega}}+2d\bigg)\triangleq K^{*}_{H}(d,M).	
\end{align}
It can be observed that the upper bound of the reconstructible sparsity level increases with $d^2$, indicating that the strength of the block-structure plays an important role in dealing with the weak-sparse scenario. Furthermore,
\begin{align} 
 	\frac{\partial K^{*}_{H}(d,M)}{\partial d} =& \sqrt{\frac{M}{1-\omega}}d+1,	\label{analysis2} \\
	\frac{\partial K^{*}_{H}(d,M)}{\partial M} =&	\frac{d^2}{4\sqrt{1-\omega}}\frac{1}{\sqrt{M}}. \label{analysis3}
\end{align}
The partial derivative of $K^{*}_{H}(d,M)$ increases proportionally with $d$ but by contrast, as $M$ increases, the corresponding partial derivative approaches 0. This means that increasing the block length $d$ always maintains an effective countermeasure against the weak-sparse problem.

Based on \textbf{Theorem~\ref{theoremmain1}}, we provide a sufficient condition for the S-BOMP/S-BOLS/FS-BOLS algorithms to achieve reliable recovery.

\begin{theoremmaindescription2}\label{theorem2}
Suppose that $\forall e\in\{1,2,\cdots,E\}$, $\|\mathbf{N}_e\|_2\leq \epsilon$, and $\mathbf{\Omega}_B^{\star}$ is the index set with $|\mathbf{\Omega}_B^{\star}|=k$. As $\mu_B,\nu\rightarrow 0$, $\forall j\in\{1,2,\cdots,k\}$, if
\begin{align}\label{theo2main2} 
	\|\mathbf{X}_{<j>}\|_F > \frac{2(k-j)\sqrt{d}\nu}{1-\nu-2j\mu_B-\sqrt{d}\nu}\|\mathbf{X}_{<j+1>}\|_F  
	 + \frac{2\sqrt{dE}\epsilon+(1+d\nu)\sum_{i\in\bar{\mathbf{\Omega}}_B^{\star}}\|\mathbf{X}_{<i>}\|_F}{\sqrt{d}(1-\sqrt{d}\nu)},
\end{align}
the S-BOMP/S-BOLS/FS-BOLS algorithms select $k$ supports from $\mathbf{\Omega}_B^{\star}$ during $k$ iterations.
\end{theoremmaindescription2}

\begin{IEEEproof}
See Appendix~\ref{proofoftheorem2}.
\end{IEEEproof}

Note that the aforementioned results apply to the BOMP and BOLS algorithms by setting the data volume $E=1$. In general, \textbf{Theorem~\ref{theorem2}} provides a sufficient condition for the S-BOMP/S-BOLS/FS-BOLS algorithms to select $k$ correct supports in $k$ iterations when facing decaying signals. Specifically, it indicates that if the signal amplitude decays in the manner described in (\ref{theo2main2}), the reliable reconstruction is guaranteed.

Consider a block orthogonal measurement matrix. The condition in \textbf{Theorem~\ref{theorem2}} become that: if
\begin{equation}\label{theo2main2varies} 
	\|\mathbf{X}_{<j>}\|_F > 2\sqrt{E}\epsilon+\frac{1}{\sqrt{d}}\sum_{i\in\bar{\mathbf{\Omega}}_B^{\star}}\|\mathbf{X}_{<i>}\|_F,
\end{equation}
the algorithms select all the $k$ correct supports. Suppose that the entries in $\mathbf{X}$ are independently and identically distributed (i.i.d.) with the Gaussian distribution $\mathcal{N}(0,\delta^2)$. Then, based on \cite[Lemma~5.1]{10}, we have
\begin{align}\label{theo2main2varies22} 
	& \Pr\Bigg(\sum_{i\in\bar{\mathbf{\Omega}}_B^{\star}}\|\mathbf{X}_{<i>}\|_F \leq \delta(\frac{N}{d}-k)\sqrt{d E+2\sqrt{d E\log (d E)}}\Bigg) \nonumber \\
	& \geq \prod_{i\in\bar{\mathbf{\Omega}}_B^{\star}}\Pr\Bigg(\|\mathbf{X}_{<i>}\|_F\leq \delta\sqrt{d E+2\sqrt{d E\log (d E)}}\Bigg) \nonumber \\
	& \geq \Big(1-\frac{1}{d E}\Big)^{\frac{N}{d}-k} ,
\end{align}
where $\Pr(\cdot)$ denotes the probability operation. Based on (\ref{theo2main2varies22}), the  condition in (\ref{theo2main2varies}) is transformed into that: $\forall j\in\{1,2,\cdots,k\}$, if 
\begin{equation}\label{theo2main2varies33} 
	\|\mathbf{X}_{<j>}\|_F > 2\sqrt{E}\epsilon + \delta\frac{2}{\sqrt{d}}\Big(\frac{N}{d}\! -\! k\Big)\sqrt{d E\! +\! 2\sqrt{d E\log (dE)}},
\end{equation}
the algorithms perform reliable recovery in $k$ iterations with the probability at least $(1-\frac{1}{d E})^{\frac{N}{d}-k}$. It can be seen that the right-side of (\ref{theo2main2varies33}) decreases with the increase of the block-length $d$, and hence the probability is also improved as $d$ increases. In other words, the stronger the block-structure characteristics, the more reliable the reconstruction. Moreover, there is a trade-off on the setting of the data volume $E$. Specifically, a larger $E$ increases the establishment probability of (\ref{theo2main2varies33}) but leads to a worse lower bound of $\|\mathbf{X}_{<j>}\|_F$. This phenomenon demonstrates the importance of choosing an appropriate $E$, which paves the way for the theoretical analysis in the subsequent section. 

\section{Data Volume Analysis}\label{datavolume} 

This section aims to reveal performance guarantees with respect to the data volume from the MIP perspective. It serves as a decision-making for choosing the necessary number of measurement vectors based on the MIP of the measurement matrix before the recovery procedure. 

\subsection{Volume Bounds for the Current Iteration}\label{currentbounds} 

The following lemma provides a sufficient condition for correct support selection by the S-BOMP/S-BOLS/FS-BOLS algorithms based on the exact recovery condition \cite{62,4}.

\begin{lemma8}\label{lemma8} 
Consider the system model (\ref{CSmodel}). Suppose that $\mathbf{X}\in\mathbb{R}^{N\times E}$ is the joint block-sparse matrix with the row-block submatrices $\mathbf{X}_{<i>}\in\mathbb{R}^{d\times E}$, $1\leq i\leq N/d=N_B$, $\mathbf{\Omega}_B^{\star}$ is the index set with $|\mathbf{\Omega}_B^{\star}|=k$, and the S-BOMP/S-BOLS/FS-BOLS algorithms have selected supports in $\mathbf{\Omega}_B^l\subseteq\mathbf{\Omega}_B^{\star}$ during the first $l$ $(1\leq l<k)$ iterations. For the $(l+1)$-th iteration, if $(d-1)\nu-(k-1)d\mu_B<1$ and the remaining row-block submatrices satisfy
\begin{align}\label{lemma8main1} 
	 \sqrt{\sum_{i\in\mathbf{\Omega}_B^{\star}\backslash\mathbf{\Omega}_B^l}\|\mathbf{X}_{<i>}\|^2_{F}}  
	  > 2\frac{1-(d-1)\nu-(k-1)d\mu_B}{\alpha_l(1-(d-1)\nu-(2k-1)d\mu_B)}\|\mathbf{N}_{i,\mathbf{\Omega}^l}\|_F,
\end{align}
the S-BOMP/S-BOLS/FS-BOLS algorithms select a correct support, where $\mathbf{N}_{i,\mathbf{\Omega}^l}=\tilde{\mathbf{C}}^T_{[i]}\mathbf{P}^\bot_{\mathbf{D}_{\mathbf{\Omega}^l}}\mathbf{N}$ and $\alpha_l$ is given in (\ref{lemma6main4}).
\end{lemma8}

\begin{IEEEproof}
See Appendix~\ref{proofoflemma8}.
\end{IEEEproof}

Different from \textbf{Lemma~\ref{lemma4}}, \textbf{Lemma~\ref{lemma8}} reveals that if the remaining power of the row-block submatrices is large enough, then the algorithms select a correct support in the $(l+1)$-th iteration. The main difference lies in the MIP related precondition. The precondition $\nu+2(l-1)\mu_B<1$ $(1\leq l<k)$ in \textbf{Lemma~\ref{lemma4}} is more relaxed than that of \textbf{Lemma~\ref{lemma8}} due to the decaying structure condition (\ref{lemma4main2}). Essentially, the both lemmas guarantee reliable support selection from the perspective of the signal power. Based on \textbf{Lemma~\ref{lemma8}}, we  present an upper data volume bound for the S-BOMP/S-BOLS/FS-BOLS algorithms to ensure correct support selection.

\begin{theoremmaindescription3}\label{theo3}
Suppose that $\forall e\in\{1,2,\cdots,E\}$, $\|\mathbf{N}_e\|_2\leq\epsilon$, the entries of the joint block-sparse matrix $\mathbf{X}$ are i.i.d. with the zero-mean Gaussian distribution, $\mathbf{\Omega}_B^{\star}$ is the index set with $|\mathbf{\Omega}_B^{\star}|=k$, and the S-BOMP/S-BOLS/FS-BOLS algorithms have selected supports in $\mathbf{\Omega}_B^l\subseteq\mathbf{\Omega}_B^{\star}$ during the first $l$ $(1\leq l<k)$ iterations. As $N\rightarrow \infty$, for the $(l+1)$-th iteration, if $(d-1)\nu-(k-1)d\mu_B<1$ and the data volume satisfies
\begin{equation}\label{theo3main1} 
	E \leq \bigg(\frac{\gamma_l\alpha_l(1-(d-1)\nu-(2k-1)d\mu_B)}{2\sqrt{d}\epsilon(1-(d-1)\nu-(k-1)d\mu_B)}\bigg)^2,
\end{equation}
 the S-BOMP/S-BOLS/FS-BOLS algorithms select a correct support with the probability at least $1-\frac{1}{(k-l)d E}$, where 
\begin{align}\label{theo3prof5} 
  \gamma_l = \sqrt{4(k-l)d E-2}  
	 -\sqrt{(k-l)d E+2\sqrt{(k-l)d E\log{(k-l)d E}}}.
\end{align}
\end{theoremmaindescription3}

\begin{IEEEproof}
See Appendix~\ref{proofoftheorem3}.
\end{IEEEproof}

\textbf{Theorem~\ref{theo3}} proposes a counter-intuitive upper bound of the data volume $E$ from the perspective of the noise impact, since a larger $E$ brings more useful information intuitively. However, the conclusion in \textbf{Theorem~\ref{theo3}} indicates that more data causes stronger noise interference at the same time.  

Denoting the right-side of (\ref{theo3main1}) as $E^{*}(\mu_B,\nu,k,d)$, we have 
\begin{equation}\label{anaE1} 
 	\frac{\partial E^{*}(\mu_B,\nu,k,d)}{\partial d} > 0.
\end{equation}
This means that a larger block length $d$ promotes stronger robustness of the recovery algorithm, which in turn increases the upper bound of the data volume. Moreover, increasing $d$ increases the lower bound $1-\frac{1}{(k-l)d E}$ of the probability $\Pr\Big(	\sqrt{\sum_{j\in\mathbf{\Omega}_B^{\star}\backslash\mathbf{\Omega}_B^l}\|\mathbf{X}_{<j>}\|^2_{F}}\geq  \gamma\Big)$ in (\ref{theo3prof4}).
 
We next derive the data volume from the perspective of useful information introduction. To this end, the sparse signals in the joint block-sparse matrix are regarded as separate individuals to isolate the influence of the data volume $E$, i.e., $\forall e\in\{1,2,\cdots,E\}$, $j\in\{1,2,\cdots,N_B\}$, $\|\mathbf{X}_{e[j]}\|_2\leq s$. Based on \textbf{Lemma~\ref{lemma8}}, the following theorem holds.

\begin{theoremmaindescription4}\label{theo4}
Suppose that $\forall e\in\{1,2,\cdots,E\}$ and $j\in\{1,2,\cdots,N_B\}$, $\|\mathbf{X}_{e[j]}\|_2\leq s$, the entries of the noise matrix $\mathbf{N}$ are i.i.d. with the Gaussian distribution $\mathcal{N}(0,\sigma^2)$, $\mathbf{\Omega}_B^{\star}$ is the index set with $|\mathbf{\Omega}_B^{\star}|=k$, and the S-BOMP/S-BOLS/FS-BOLS algorithms have selected supports in $\mathbf{\Omega}_B^l\subseteq\mathbf{\Omega}_B^{\star}$ during the first $l$ $(1\leq l<k)$ iterations. For the $(l+1)$-th iteration, if $(d-1)\nu-(k-1)d\mu_B<1$ and the data volume satisfies
\begin{equation}\label{theo4main11} 
	E \geq \frac{d\zeta_l^2}{(k-l)s},
\end{equation}
where
\begin{equation}\label{theo4main2} 
	\zeta_l = \sigma\sqrt{(k-l)E d+2\sqrt{(k-l)E d\log(k-l)E d}},
\end{equation} 
 the S-BOMP/S-BOLS/FS-BOLS algorithms select a correct support with the probability at least $1-\frac{1}{(k-l)d E}$.
\end{theoremmaindescription4}
	
\begin{IEEEproof}
See Appendix~\ref{proofoftheorem4}.
\end{IEEEproof}

Based on \textbf{Theorems~\ref{theo3}} and \textbf{\ref{theo4}}, the following corollary holds.

\begin{Corollary2}\label{coro2} 
Suppose that the entries of the joint block-sparse matrix and the noise matrix are i.i.d. with the Gaussian distributions $\mathcal{N}(0,1)$ and $\mathcal{N}(0,\sigma^2)$, respectively, $\mathbf{\Omega}_B^{\star}$ is the index set with $|\mathbf{\Omega}_B^{\star}|=k$, and the S-BOMP/S-BOLS/FS-BOLS algorithms have selected supports in $\mathbf{\Omega}_B^l\subseteq\mathbf{\Omega}_B^{\star}$ during the first $l$ $(1\leq l<k)$ iterations. For the $(l+1)$-th iteration, if $(d-1)\nu-(k-1)d\mu_B<1$ and the following inequality holds:
\begin{align}\label{coro2main} 
 	\gamma_l > 2\frac{1-(d-1)\nu-(k-1)d\mu_B}{\alpha_l(1-(d-1)\nu-(2k-1)d\mu_B)}  
 	 \times \sqrt{1-(d-1)\nu}\sigma\sqrt{k E d+2\sqrt{k E d\log{k E d}}},
\end{align}
the S-BOMP/S-BOLS/FS-BOLS algorithms can select a correct support with the probability at least $1-\frac{1}{(k-l)d E}-\frac{1}{k d E}$, where $\gamma_l$ is given in (\ref{theo3prof5}).
\end{Corollary2}

\begin{IEEEproof}
See Appendix~\ref{proofofcorollary1}.
\end{IEEEproof}

\textbf{Corollary~\ref{coro2}} is an integrated extension of \textbf{Theorems~\ref{theo3}} and \textbf{\ref{theo4}}. It offers a prior judgment rule if the data volume is sufficient for reliable recovery with the other parameters fixed. Similarly, when fixing $E$, (\ref{coro2main}) seems to provide decision-making for the other parameter settings, such as $k$, $d$ and $\sigma$, before joint block-sparse recovery. However, $k$, $d$ or $\sigma$ are the inherent properties of the sparse signal or the environment, and therefore they are difficult to be changed. Hence, the parameter $E$'s value setting is the core meaning of (\ref{coro2main}).

Moreover, the correctness of the support selection in the first iteration makes great sense to the algorithms' subsequent operation. Thus, based on \textbf{Corollary~\ref{coro2}}, the following corollary provides the bounds of the data volume that ensures the algorithms to correctly select a support in the first iteration under the asymptotic case.

\begin{Corollary3}\label{coro3} 
Suppose that the entries of the sparse signal matrix and the noise matrix are i.i.d. with the Gaussian distributions $\mathcal{N}(0,1)$ and $\mathcal{N}(0,\sigma^2)$, respectively. If $\mu_B,\nu\rightarrow0$ and the data volume satisfies
\begin{align}\label{coro3main} 
 	 \frac{8-2c-2c^2-\sqrt{4c^4+8c^3-16c^2}}{16-8c-3c^2}\leq E  
 	  \leq \frac{8-2c-2c^2+\sqrt{4c^4+8c^3-16c^2}}{16-8c-3c^2},
\end{align}
the S-BOMP/S-BOLS/FS-BOLS algorithms select a correct support in the first iteration with the probability at least $1-\frac{2}{k d E}$, where $c=\big(\frac{2}{\sqrt{d}}\sigma+1\big)^2$.
\end{Corollary3}

\begin{IEEEproof}
As $\mu_B,\nu\rightarrow0$, by direct simplification, the sufficient condition (\ref{coro2main}) becomes
\begin{equation}\label{remakr31} 
  \big((4-c)k d E-2\big)^2>4c^2(k d E\log (k d E)).
\end{equation}
According to the Taylor series, $\log (k d E)\leq k d E - 1$. Therefore, the sufficient condition (\ref{remakr31}) becomes
\begin{equation}\label{coro32} 
	(16-8c-3c^2)(k d E)^2+(4c^2+4c-16) k d E+4 > 0 .
\end{equation}
Note that the resolution of (\ref{coro32}) is (\ref{coro3main}). This completes the proof.
\end{IEEEproof}

Letting $\sigma=0$, we have $c=1$, and the discriminant of the quadratic equation in (\ref{coro32}) with respect to $k d E$ satisfies $\Delta=24c^3-32c^2<0$. This result indicates that $E \geq 1$ is sufficient for reliable recovery in the noiseless scenario, and at least one data signal is required for recovery.

\subsection{Overall Volume for Reliable Recovery}\label{S5.2}

The theoretical results in Subsection~\ref{currentbounds} provide guarantees of the correct support selection for the S-BOMP/S-BOLS/FS-BOLS in the $l$-th iteration $(1\leq l<k)$. By direct calculation, we have $\frac{\partial \alpha_l}{\partial l}<0$. Thus, the following inequality holds:
\begin{equation}\label{analysis222} 
	\alpha_l \geq \alpha_{k-1} = \sqrt{\frac{d(1-\nu-2(k-2)\mu_B-4\mu_B^2)}{1-\nu-2(k-2)\mu_B}}.
\end{equation}
Based on (\ref{analysis222}), a sufficient condition for (\ref{lemma8main1}) to hold is
\begin{equation}\label{lemma9main11} 
	\sqrt{\sum_{j\in\mathbf{\Omega}_B^{\star}\backslash\mathbf{\Omega}_B^l}\|\mathbf{X}_{<j>}\|^2_{F}} > G(\mu_B,\nu,k,d)\|\mathbf{N}_{i,\mathbf{\Omega}}\|_F,
\end{equation}
where 
\begin{align}\label{lemma9main22} 
  G(\mu_B,\nu,k,d) = 2\frac{1-(d-1)\nu-(k-1)d\mu_B}{1-(d-1)\nu-(2k-1)d\mu_B}  
   \times \sqrt{\frac{1-\nu-2(k-2)\mu_B}{d(1-\nu-2(k-2)\mu_B-4\mu_B^2)}}.
\end{align}

As usual, we compare our result in (\ref{lemma9main11}) with the existing one in the literature. In \cite{3}, the authors provide a condition for the reliable recovery of the SMV model as
\begin{equation}\label{smvm1} 
	\|\mathbf{x}_{\mathbf{\Omega}^{\star}\backslash\mathbf{\Omega}}\|_{2}>\frac{2\sqrt{K}}{1-(2K-1)\mu}\|\mathbf{n}_{i,\mathbf{\Omega}}\|_2,
\end{equation}
where $\mathbf{x}$ and $\mathbf{n}$ are the corresponding variants of $\mathbf{X}$ and $\mathbf{N}$ in vector form. Correspondingly, the condition in (\ref{lemma9main11}) for the SMV model is directly given by 
\begin{equation}\label{smvm2} 
	\|\mathbf{x}_{\mathbf{\Omega}^{\star}\backslash\mathbf{\Omega}}\|_{2}>G(\mu_B,\nu,k,d)\|\mathbf{n}_{i,\mathbf{\Omega}}\|_2.
\end{equation}
It can be observed that $G(\mu_B,\nu,k,d)$ acts as a pivotal function for the establishment of (\ref{smvm2}). Since $\nu,\mu_B\leq\mu$ and for the case that $d\geq2$, we have $G(\mu_B,\nu,k,d)\leq\frac{2\sqrt{K}}{1-(2K-1)\mu}$, which reveals that the condition in (\ref{smvm2}) is more relaxed than that in (\ref{smvm1}). This result effectively reduces the required power of the remaining support blocks for reliable recovery. Furthermore, letting $d=1$, the condition in (\ref{smvm2}) becomes
\begin{align}\label{smvm3} 
	\|\mathbf{x}_{\mathbf{\Omega}^{\star}\backslash\mathbf{\Omega}}\|_{2} > 2\frac{1-(K-1)\mu}{1-(2K-1)\mu}  
	 \times \sqrt{\frac{1-2(K-2)\mu}{(1-2(K-2)\mu-4\mu^2)}}\|\mathbf{n}_{i,\mathbf{\Omega}}\|_2.
\end{align}
Similar conclusion that (\ref{smvm3}) is more relaxed than (\ref{smvm1}) can be obtained. 

Based on the aforementioned analysis, the following corollary presents a condition for the algorithms to select all the correct supports in $k$ iterations.

\begin{Corollary4}\label{lemma10} 
Suppose that $\forall e\in\{1,2,\cdots,E\}$, $\|\mathbf{N}_e\|_2\leq\epsilon$, $\mathbf{X}\in\mathbb{R}^{N\times E}$ is the joint block-sparse matrix with the row-block submatrices $\mathbf{X}_{<i>}\in\mathbb{R}^{d\times E}$ $(1\leq i\leq N/d=N_B)$, and $\mathbf{\Omega}_B^{\star}$ is the index set with $|\mathbf{\Omega}_B^{\star}|=k$. 
	If $(d-1)\nu-(k-1)d\mu_B<1$, and $\forall j\in\mathbf{\Omega}_B^{\star}$,
\begin{equation}\label{lemma9main1} 
	\|\mathbf{X}_{<j>}\|_{F}>\frac{1}{\sqrt{k}}G(\mu_B,\nu,k,d)\sqrt{d E}\epsilon,
\end{equation}
the S-BOMP/S-BOLS/FS-BOLS algorithms select the $k$ supports from $\mathbf{\Omega}_B^{\star}$ during $k$ iterations, where 
\begin{align}\label{lemma9main2} 
	G(\mu_B,\nu,k,d) = 2\frac{1-(d-1)\nu-(k-1)d\mu_B}{1-(d-1)\nu-(2k-1)d\mu_B}  
	 \times \sqrt{\frac{1-\nu-2(k-2)\mu_B}{d(1-\nu-2(k-2)\mu_B-4\mu_B^2)}}.
\end{align}
\end{Corollary4}

The proof is straightforward and therefore is omitted. In particular, (\ref{lemma9main1}) is a sufficient condition of (\ref{lemma9main11}) under the assumption $\|\mathbf{N}_e\|_2\leq\epsilon$ $(\forall e\in\{1,2,\cdots,E\})$. 

Similarly, since $\gamma_l$ defined in \textbf{Theorem~\ref{theo3}} satisfies $\frac{\partial\gamma_l}{\partial l}<0$, and $\zeta_l$ given in \textbf{Theorem~\ref{theo4}} satisfies $\frac{\partial\zeta_l}{\partial l}<0$, we have
\begin{align} 
	\gamma_l \geq& \gamma_{k-1} = \sqrt{4 d E-2} - \sqrt{d E+2\sqrt{d E\log(d E)}}, \label{theo3prof6} \\
	\zeta_l \leq& \zeta_0 = \sigma\sqrt{k E d+2\sqrt{k E d\log(k E d)}}. \label{theo4main22}
\end{align}
Based on (\ref{analysis222}), (\ref{theo3prof6}) and (\ref{theo4main22}), \textbf{Theorems~\ref{theo3}} and \textbf{\ref{theo4}} can be extended to the following two corollaries, respectively.

\begin{Corollary4}\label{coro4}
Suppose that $\forall e\in\{1,2,\cdots,E\}$, $\|\mathbf{N}_e\|_2\leq\epsilon$, $\mathbf{\Omega}_B^{\star}$ is the index set with $|\mathbf{\Omega}_B^{\star}|=k$, and the entries of the sparse matrix $\mathbf{X}$ are i.i.d. with the Gaussian distribution. As $N\rightarrow\infty$, $\forall 1\leq l<k$, if $(d-1)\nu-(k-1)d\mu_B<1$ and the data volume satisfies
\begin{equation}\label{coro4main1} 
	E \leq \bigg(\frac{\gamma_{k-1}\alpha_{k-1}(1-(d-1)\nu-(2k-1)d\mu_B)}{2\sqrt{d}\epsilon(1-(d-1)\nu-(k-1)d\mu_B)}\bigg)^2,
\end{equation}
the S-BOMP/S-BOLS/FS-BOLS algorithms select $k$ supports from the set $\mathbf{\Omega}_B^{\star}$ during $k$ iterations with the probability at least $1-\frac{1}{dE}$, where $\gamma_{k-1}$ and $\alpha_{k-1}$ are given in (\ref{analysis222}) and (\ref{theo3prof6}), respectively.
\end{Corollary4}	

\begin{Corollary5}\label{coro5}
Suppose that $\forall e\in\{1,2,\cdots,E\}$, $j\in\{1,2,\cdots,N_B\}$, $||\mathbf{x}_{e[j]}||_2\leq s$, $\mathbf{\Omega}_B^{\star}$ is the index set with $|\mathbf{\Omega}_B^{\star}|=k$, and the entries of the noise matrix $\mathbf{N}$ are i.i.d. with the Gaussian distribution $\mathcal{N}(0,\sigma^2)$. $\forall l<k$, if $(d-1)\nu-(k-1)d\mu_B<1$ and the data volume satisfies
\begin{equation}\label{theo4main1} 
	E \geq \frac{d\zeta_0^2}{s},
\end{equation}
the S-BOMP/S-BOLS/FS-BOLS algorithms select $k$ supports from $\mathbf{\Omega}_B^{\star}$ in $k$ iterations with probability at least $1-\frac{1}{kdE}$, where $\zeta_0$ is defined in (\ref{theo4main22}).
\end{Corollary5}	

As \textbf{Corollaries~\ref{coro4}} and \textbf{\ref{coro5}} provide universal conditions of the data volume to perform reliable recovery, they are more strict than the results in \textbf{Theorems~\ref{theo3}} and \textbf{\ref{theo4}}, but offer more concise and useful bounds. Specifically, the decision-makings of the data volume $E$ based on \textbf{Corollaries~\ref{coro4}} and \textbf{\ref{coro5}} only need to be checked once before reconstruction, while those of \textbf{Theorems~\ref{theo3}} and \textbf{\ref{theo4}} need to be verified every iteration.

We now present the following theorem to give upper and lower bounds of the data volume in closed forms, based on the widely-used assumption that the nonzero entries of the signal and the noise are i.i.d. with Gaussian distributions. 

\begin{theoremmaindescription5}\label{theo5}
Suppose that the entries of the sparse signal matrix and the noise matrix are i.i.d. with the Gaussian distributions $\mathcal{N}(0,\xi^2)$ and $\mathcal{N}(0,\sigma^2)$, respectively, $\mathbf{\Omega}_B^{\star}$ is the index set with $|\mathbf{\Omega}_B^{\star}|=k$, and $k\le d$. If $(d-1)\nu-(k-1)d\mu_B<1$ and the data volume satisfies
\begin{equation}\label{theo5main} 
	\max\{1,\underline{E}(\mu_B,\nu,\xi,\sigma,d)\}\leq E\leq\overline{E}(\mu_B,\nu,\xi,\sigma,d),
\end{equation}
the S-BOMP/S-BOLS/FS-BOLS algorithms select $k$ supports from $\mathbf{\Omega}_B^{\star}$ during $k$ iterations with the probability at least $1-\frac{1}{d E}-\frac{1}{k d E}$. 

In (\ref{theo5main}), $\underline{E}(\mu_B,\nu,\xi,\sigma,d)=\frac{-b_1-\sqrt{\Delta}}{2a_1}$, $\overline{E}(\mu_B,\nu,\xi,\sigma,d)=\frac{-b_1+\sqrt{\Delta}}{2a_1}$,
\begin{align} 
 a_1 =& (16\xi^4-8J\xi^2-3J^2)d^2 , \label{eq-a1} \\
 b_1 =& (-16\xi^4+4J\xi^2+4J^2)d , \label{eq-b1} \\
 c_1 =& 4\xi^4, \label{eq-c1}
\end{align}
 $J=(\mathcal{T}\sigma+\xi)^2$, $\Delta=b_1^2-4a_1c_1$ and 
\begin{align}\label{theo555prof2} 
	\mathcal{T} =& 2\frac{1-(d-1)\nu-(k-1)d\mu_B}{1-(d-1)\nu-(2k-1)d\mu_B}  
	 \times \sqrt{\frac{(1-\nu-2(k-2)\mu_B)(1-(d-1)\nu)}{1-\nu-2(k-2)\mu_B-4\mu_B^2}}.
\end{align}
\end{theoremmaindescription5}

\begin{IEEEproof}
See Appendix~\ref{proofoftheorem5}.
\end{IEEEproof}

Based on \textbf{Theorem~\ref{theo5}}, we exploit asymptotic analysis to further explore the role of block-structure characteristic in operating the data volume in the following corollary.

\begin{Corollary6}\label{Corollary6}
Suppose that the entries of the sparse signal matrix and the noise matrix are i.i.d. with the Gaussian distributions $\mathcal{N}(0,\xi^2)$ and $\mathcal{N}(0,\sigma^2)$, respectively, $\mathbf{\Omega}_B^{\star}$ is the index set with $|\mathbf{\Omega}_B^{\star}|=k$, and $k\le d$.	If $\mu_B,\nu\rightarrow0$, and the data volume satisfies
\begin{equation}\label{Corollary6main} 
	\max\{1,\underline{E}^*(\xi,\sigma,d)\}\leq E\leq\overline{E}^*(\xi,\sigma,d),
\end{equation}
the S-BOMP/S-BOLS/FS-BOLS algorithms select $k$ supports from $\mathbf{\Omega}_B^{\star}$ during $k$ iterations with the probability at least $1-\frac{1}{dE}-\frac{1}{kdE}$, where $\underline{E}^*(\xi,\sigma,d)=\frac{-b_2-\sqrt{\Delta}}{2a_2}$, $\overline{E}^*(\xi,\sigma,d)=\frac{-b_2+\sqrt{\Delta}}{2a_2}$, $a_2=(16\xi^4-8J\xi^2-3J^2)d^2$, $b_2=(-16\xi^4+4J\xi^2+4J^{2})d$, $c_2=4\xi^4$, $J=(2\sigma+\xi)^2$ and $\Delta=b_2^2-4a_2c_2$.
\end{Corollary6}

The proof is omitted since it is similar to that of \textbf{Theorem~\ref{theo5}}.

For $\underline{E}^*(\xi,\sigma,d)$ in the lower bound of (\ref{Corollary6main}), we have $\lim\limits_{d\rightarrow\infty}\underline{E}^*(\xi,\sigma,d)=0$. As presented in \textbf{Theorem~\ref{theo4}}, the lower bound of the data volume is regarded as the quantity of the input of useful information. As $\underline{E}^*(\xi,\sigma,d)$ approaches 0, the required number of useful signals are decreasing accordingly. This reveals that large block length $d$ induces strong power in reliable recovery, leading to lower data volume required for reconstruction. The upper bound $\overline{E}^*(\xi,\sigma,d)$ in (\ref{Corollary6main}) also decreases as the block length $d$ increases under the asymptotic assumption of \textbf{Theorem~\ref{theo5}}. The reason of this phenomenon is that as the block length $d$ increases, the total sparsity of the sparse signal is also increased, which poses huge challenge to the reliable recovery. It follows that a lower sparsity induces better performance as stated in the reconstructible sparsity analysis in Subsection~\ref{mainconditons}. Thus, combining the analyses of $\underline{E}^*(\xi,\sigma,d)$ and $\overline{E}^*(\xi,\sigma,d)$, it is concluded that a larger $d$ can produce reliable gain for recovery when the overall sparsity is fixed but may cause performance degradation if the overall sparsity is increased due to the larger $d$ under the assumption of \textbf{Corollary~\ref{Corollary6}}. 

Moreover, similar to the discussions of \textbf{Corollary~\ref{coro3}}, setting $\sigma=0$, the discriminant of corresponding quadratic equation satisfies $\Delta<0$ and the volume bound in (\ref{theo5main}) changes into $E\geq 1$. This is consistent with the intuition that at least one measurement vector is required for recovery. Since \textbf{Theorem~\ref{theo5}} is a sufficient condition for the correct selection of all the $k$ supports, $E\geq 1$ indicates that one measurement vector is sufficient for reliable reconstruction in the noiseless scenario.
	
\section{Conclusions}\label{S6}

The focus of this paper has been to reveal the recovery conditions of the joint block-sparse problem. Specifically, we have proposed the two OLS framework-based S-BOLS and FS-BOLS algorithms and have provided the theoretical analysis for them in parallel with the S-BOMP algorithm. Our derived results have indicated that under certain conditions combined with MIP and the gradient of signal decaying, the S-BOMP/S-BOLS/FS-BOLS algorithms perform reliable recovery during $k$ iterations. Correspondingly, we have presented an improved upper bound of the reconstructible sparsity, which offers a sufficient condition of the exact recovery condition under the decaying signal and noiseless case, and it implies that the requirement on the sparsity level becomes more relaxed as the block length increases. We have also derived the sufficient conditions of the data volume from the perspective of useful information and noise impact, which provide the lower and upper bounds of the data volume, respectively. The derived upper bound of the data volume, which has not been analyzed for most of the joint recovery methods based on the MIP technique, seems to be counter-intuitive at the first glance, but it actually makes sense as a larger number of the measurement vectors introduce more noise impact and thus the algorithms cannot perform reliable reconstruction. Asymptotic analyses have further been derived to clearly show  the essential relationships between the recovery conditions with the matrix coherence, the number of measurements, the block sparsity and the block-structure. Overall, the smaller the matrix coherence and the block sparsity, the better the theoretical conditions, and the converse results correspond to the number of measurements and the block-structure. Theoretical comparisons have confirmed superiority of our developed results over the existing results in the literature,  in terms of performance reliability of the related algorithms. 

\appendix

\subsection{Proof of \textbf{Lemma~\ref{lemma1}}}\label{proofoflemma1} 

\begin{IEEEproof}	
For the lower bound of $\lambda_{\min}$, we prove that the matrix $\mathbf{D}^{\rm T}\mathbf{D}-\lambda \mathbf{I}$ is nonsingular under the condition $\lambda < 1-(d-1)\nu-(k-1)d\mu_B$ when $(d-1)\nu+(k-1)d\mu_B < 1$. This is equivalent to prove that for any nonzero vector $\mathbf{f}=[\mathbf{f}_1,\mathbf{f}_2,\cdots,\mathbf{f}_{kd}]^{\rm T}\in \mathbb{R}^{kd}$, $(\mathbf{D}^{\rm T}\mathbf{D}-\lambda \mathbf{I})\mathbf{f}\neq \mathbf{0}$. Without loss of generality, we assume $\|\mathbf{f}_{[1]}\|_2\geq\|\mathbf{f}_{[2]}\|_2\geq \cdots\geq \|\mathbf{f}_{[k]}\|_2$. The $\ell_2$-norm of the first row-block of $(\mathbf{D}^{\rm T}\mathbf{D}-\lambda \mathbf{I})\mathbf{f}$ satisfies
\begin{align}\label{thefirstblock1} 
	& \big\|\big[(\mathbf{D}^{\rm T}\mathbf{D}-\lambda \mathbf{I})\mathbf{f}\big]_{[1]}\big\|_2 \nonumber \\
	& = \bigg\|\big(\mathbf{D}_{[1]}^{\rm T}\mathbf{D}_{[1]}-\lambda \mathbf{I}\big)\mathbf{f}_{[1]} + \sum_{i=2}^{k} \mathbf{D}_{[1]}^{\rm T}\mathbf{D}_{[i]}\mathbf{f}_{[i]}\bigg\|_2 \nonumber \\
	& \geq \big\|\mathbf{D}_{[1]}^{\rm T}\mathbf{D}_{[1]}\mathbf{f}_{[1]}\big\|_2 - \lambda \| \mathbf{f}_{[1]}\|_2 - d \mu_B \sum_{i=2}^k\|\mathbf{f}_{[i]}\|_2.
\end{align}
Denote $\mathbf{D}_{[1]}^{\rm T}\mathbf{D}_{[1]}=\mathbf{I}+\mathbf{B}$, where $\forall i$, $\mathbf{B}_{ii}=0$. Based on Ger\v{s}gorin's disc theorem \cite{4} 
\begin{align}\label{thefirstblock2} 
	\|\mathbf{B}\|_2\leq (d-1)\nu.
\end{align}
Thus, 
\begin{align}\label{thefirstblock3} 
	\big\|\mathbf{D}_{[1]}^{\rm T}\mathbf{D}_{[1]}\big\|_2\geq \|\mathbf{I}\|_2-\|\mathbf{B}\|_2\geq 1-(d-1)\nu .
\end{align}
Combining (\ref{thefirstblock1}) and (\ref{thefirstblock3}) leads to
\begin{align}\label{thefirstblock4} 
	& \big\|\big[(\mathbf{D}^{\rm T}\mathbf{D}-\lambda \mathbf{I})\mathbf{f}\big]_{[1]}\big\|_2 \nonumber \\	
  & \geq (1-(d-1)\nu-\lambda )\|\mathbf{f}_{[1]}\|_2 - d\mu_B \sum_{i=2}^k\|\mathbf{f}_{[i]}\|_2 \nonumber \\
	& > (k-1)d \mu_B\|\mathbf{f}_{[1]}\|_2 - d \mu_B \sum_{i=2}^k\|\mathbf{f}_{[i]}\|_2 \geq 0 .
\end{align}
This reveals that $(\mathbf{D}^{\rm T}\mathbf{D}-\lambda \mathbf{I})\mathbf{f}\neq \mathbf{0}$ and $\lambda_{\min}\geq 1-(d-1)\nu-(k-1)d\mu_B$ holds. 

The inequality $\lambda_{\max}\leq1+(d-1)\nu+(k-1)d\mu_B$ can be proved similarly.
\end{IEEEproof}

\subsection{Proof of \textbf{Proposition~\ref{Props2}}}\label{proofofprops2} 

\begin{IEEEproof}
We need to derive the condition for the gap between $F^{l}$ and $F^{l}_r$ to be sufficiently small based on the conventional coherence and block-structure coherence, respectively. 
 	
\noindent
$\bullet$ Proof for the non block-structure measurement matrix:
 
The first terms inside the maximization operators of (\ref{ctri}) and (\ref{new-ctri}) are equal. By dividing the second term in (\ref{ctri}) by the second term in (\ref{new-ctri}), we have
\begin{equation}\label{divide} 
 	\mathcal{G} =\Bigg(\frac{\|\mathbf{P}^\bot_{\mathbf{D}_{\mathbf{\Theta}^{l}}}\mathbf{D}_{(j_B-1)d+2}\|_2}{\|\mathbf{P}^\bot_{\mathbf{D}_{\mathbf{\Theta}^{l}\cup\{(j_B-1)d+1\}}}\mathbf{D}_j\|_2}\Bigg)^2.
\end{equation}
It is proved in \cite{8} that when $(K-1)\mu<1$, $\sqrt{1-(K-1)\mu}\leq \|\mathbf{P}^{\bot}_{\mathbf{D}_{\mathbf{\Theta}^l}}\mathbf{D}_i\|_2\leq1$, where $K-1$ is the largest cardinality of the set $\mathbf{\Theta}^l$. Thus, (\ref{divide}) is bounded by
 \begin{equation}	\label{bounded} 
	\begin{aligned}
		1-(K-1)\mu\leq\mathcal{G}\leq\frac{1}{1-(K-1)\mu}.
	\end{aligned}
\end{equation}
Based on the squeeze theorem \cite{squeeze-theorem}, we obtain 
\begin{equation}\label{probis2} 
	\lim_{\mu\rightarrow0}\mathcal{G}=1.
\end{equation}
Similar results can be obtained for the other term pairs in (\ref{ctri}) and (\ref{new-ctri}). 

The result (\ref{probis2}) for all these term pairs then reveals that when $\mu\rightarrow0$, the block indices selected by the rules (\ref{ctri}) and (\ref{new-ctri}) are the same. For the Gaussian measurement matrix, its reachable lower bound of $\mu$ is $\sqrt{\frac{N-M}{M(N-1)}}$ \cite{58}. In the asymptotic case, when $\omega$ is fixed,
\begin{equation}\label{nmrelation} 
  \lim_{M/N=\omega;\;M,N\rightarrow\infty}\sqrt{\frac{N-M}{M(N-1)}}=\sqrt{\frac{1-\omega}{M}}=0,
\end{equation}
which indicates that $\lim\limits_{M/N=\omega;\;M,N\rightarrow\infty}\mu=0$.

\noindent	$\bullet$ Proof for the block-structure measurement matrix:

Now consider a block-structure measurement matrix. Based on \textbf{Lemma~\ref{lemma1}} and \cite[Lemma 5]{3}, when $(d-1)\nu+(k-1)d\mu_B<~1$, we have
\begin{equation}\label{block-bounds} 
	\sqrt{1-(d-1)\nu-(k-1)d\mu_B}\leq \|\mathbf{P}^{\bot}_{\mathbf{D}_{\mathbf{\Theta}^l}}\mathbf{D}_i\|_2\leq 1,  
\end{equation}
since the largest cardinality of the set $\mathbf{\Theta}_B^l$ is $k-1$, where $\mathbf{\Theta}_B^l$ is the corresponding block index set of $\mathbf{\Theta}^l$. Then, $\mathcal{G}$ is further bounded by
\begin{equation}\label{further_bounded} 
	\begin{aligned}
		1-(d-1)\nu-  (k-1)d\mu_B\leq \mathcal{G} 
		 \leq \frac{1}{1-(d-1)\nu-(k-1)d\mu_B} .
	\end{aligned}
\end{equation}
It can be observed that 
\begin{equation}\label{probis3} 
	\lim_{\mu_B\rightarrow 0,\nu\rightarrow0} \mathcal{G} = 1 .
\end{equation}
Meanwhile, since $\mu_B,\nu\in[0,~ \mu]$ \cite{1}, the attainable lower bounds of $\mu_B$ and $\nu$, denoted as $\mu_{B\min}$ and $\nu_{\min}$, in the case of Gaussian measurement matrix satisfy 
\begin{equation}	\label{muhenu} 
	0 \leq \mu_{B\min}, \nu_{\min}\leq \sqrt{\frac{N-M}{M(N-1)}}.
\end{equation}
The squeeze  theorem implies that
\begin{equation}\label{probis4} 
	\lim_{M/N=\omega;\;M,N\rightarrow\infty}\mu_{B\min}, \nu_{\min}=0.
\end{equation}

This completes the proof of \textbf{Proposition~\ref{Props2}}.
\end{IEEEproof}

\subsection{Proof of \textbf{Lemma~\ref{lemma3}}}\label{proofoflemma3} 

\begin{IEEEproof}
Define $\mathbf{U}=\mathbf{D}^{\rm T}_{\mathbf{\Theta}}\mathbf{D}_{\mathbf{\Psi}}\in\mathbb{R}^{g d\times h d}$. It is noted that there exist $g h$ blocks of size $d\times d$ in $\mathbf{U}$. Express $\mathbf{U}_{[i,j]}$ as the $(i,j)$-th $d\times d$ block submatrix in $\mathbf{U}$. Then we have
\begin{align} 
	\|\mathbf{U}\|_2 & \leq  \min\{\|\mathbf{U}\|_1,\|\mathbf{U}\|_{\infty}\} \label{u1} \\
	\leq & \min\Bigg\{\sum_{i=1}^{g}\max_{j\in\{1,2,\cdots,h\}}\|\mathbf{U}_{[i,j]}\|_1 ,  
	 \sum_{j=1}^{h}\max_{i\in\{1,2,\cdots,g\}}\|\mathbf{U}_{[i,j]}\|_{\infty}\Bigg\} \label{u2} \\
	\leq & \min\Bigg\{\sum_{i=1}^{g}\max_{j\in\{1,2,\cdots,h\}}\sqrt{d}\|\mathbf{U}_{[i,j]}\|_2 ,  
	 \sum_{j=1}^{h}\max_{i\in\{1,2,\cdots,g\}}\sqrt{d}\|\mathbf{U}_{[i,j]}\|_{2}\Bigg\} \nonumber \\
	\leq & \min\Big\{g d^{\frac{3}{2}}\mu_B, h d^{\frac{3}{2}}\mu_B\Big\} , \label{u3}
\end{align}
where the inequality (\ref{u1}) follows from Ger\v{s}gorin's disc theorem, (\ref{u2}) is based on the relationship between $\mathbf{U}$ and its block submatrices, and (\ref{u3}) follows from the definition of block-coherence. 
\end{IEEEproof}

\subsection{Proof of \textbf{Lemma~\ref{lemma2}}}\label{proofoflemma2} 

\begin{IEEEproof}
Note that 
\begin{align}	
 	\|\mathbf{P}^\bot_{\mathbf{D}_{\mathbf{\Theta}}}\mathbf{D}_{\mathbf{\Psi}}\mathbf{x}_{\mathbf{\Psi}}\|_2^2 & =\|\mathbf{D}_{\mathbf{\Psi}}\mathbf{x}_{\mathbf{\Psi}}\|_2^2 - \|\mathbf{P}_{\mathbf{D}_{\mathbf{\Theta}}}\mathbf{D}_{\mathbf{\Psi}}\mathbf{x}_{\mathbf{\Psi}}\|_2^2 \nonumber \\
  \geq & (1-\underline{\delta}_{h d,0})\|\mathbf{x}_{\mathbf{\Psi}}\|^2_2 - \frac{\|\mathbf{D}^{\rm T}_{\mathbf{\Theta}} \mathbf{D}_{\mathbf{\Psi}} \mathbf{x}_{\mathbf{\Psi}}\|_2^2}{1 - \underline{\delta}_{g d,0}} \label{aaa1} \\
 	\geq & (1-\underline{\delta}_{h d,0})\|\mathbf{x}_{\mathbf{\Psi}}\|^2_2  
 	 - \frac{\min\big\{g^2d^3\mu_B^2, h^2d^{3}\mu_B^2\big\} \|\mathbf{x}_{\mathbf{\Psi}}\|_2^2}{1-\underline{\delta}_{g d,0}}, \label{aaa2}
\end{align}	 
where (\ref{aaa1}) follows from the relationships between the RICs of $\mathbf{D}$ and the transforms of $\mathbf{D}$ (see \cite[Lemma 4]{60}), and (\ref{aaa2}) is derived based on \textbf{Lemma~\ref{lemma3}}. Then, (\ref{aaa2}) unveils (\ref{anylowerP-RIP}).
	
Based on (\ref{conventionalRIP}) and \textbf{Lemma~\ref{lemma1}}, we have
\begin{align}\label{lemma2proof1} 
  \|\mathbf{P}^\bot_{\mathbf{D}_{\mathbf{\Theta}}}\mathbf{D}_{\mathbf{\Psi}}\mathbf{x}_{\mathbf{\Psi}}\|_2^2 & =\|\tilde{\mathbf{D}}^{\mathbf{\Theta}}_{\mathbf{\Psi}}\mathbf{x}_{\mathbf{\Psi}}\|_2^2 \leq \|\mathbf{D}_{\mathbf{\Psi}}\mathbf{x}_{\mathbf{\Psi}}\|_2^2  
 	\leq  (1+(d-1)\nu+(h-1)d\mu_B)\|\mathbf{x}_{\mathbf{\Psi}}\|_2^2 ,
\end{align}
which indicates that $\bar{\delta}_{h,0}=(d-1)\nu+(h-1)d\mu_B$ and (\ref{anyhigherP-RIP}) is thus a consequence. This completes the proof.	
\end{IEEEproof}

\subsection{Proof of \textbf{Lemma~\ref{lemma6}}}\label{proofoflemma6} 

\begin{IEEEproof}
From \textbf{Lemma~\ref{lemma2}}, we have
\begin{align} 
	\underline{\delta}_{2,2l} =& \nu + \frac{8\mu_B^2}{1-\nu-2(l-1)\mu_B} , \label{lemma61} \\
	\bar{\delta}_{2,2l} =& \nu . \label{lemma62}
\end{align}
Set $d=1$. According to \textbf{Lemma~\ref{lemma5}}, $\forall i\neq j$,
\begin{equation}\label{lemma6proof1} 
	|\langle\tilde{\mathbf{D}}_{i},\tilde{\mathbf{D}}_{j}\rangle| \leq \Big(\nu + \frac{4\mu_B^2}{1-\nu-2(l-1)\mu_B}\Big).
\end{equation}
Similarly, $\forall i\neq j$,
\begin{equation}\label{lemma63} 
	\|\tilde{\mathbf{D}}^T_{[i]}\tilde{\mathbf{D}}_{[j]}\|_F\leq d\Big(\nu+\frac{4\mu_B^2}{1 - \nu - 2(l - 1)\mu_B}\Big).
\end{equation}
To obtain $\beta_l(\mu_B,\nu,d)$, it remains to develop a lower bound of $\|\tilde{\mathbf{D}}_{i}\|_2$, denoted by $\tau_{\min}$, since $\|\tilde{\mathbf{C}}^{\rm T}_{[i]}\tilde{\mathbf{D}}_{[j]}\|_F\leq\| \tilde{\mathbf{D}}^{\rm T}_{[i]}\tilde{\mathbf{D}}_{[j]}\|_F/\tau_{\min}$. Noting $\nu\leq \mu$, then based on \cite[Lemma 12]{60} we have
\begin{align}\label{lemma64} 
  \|\tilde{\mathbf{D}}_{i}\|_2^2 \geq & 1 + \nu-	|\langle\tilde{\mathbf{D}}_{i},\tilde{\mathbf{D}}_{j}\rangle| \nonumber \\
	\geq &\frac{1-\nu-2(l-1)\mu_B-4\mu_B^2}{1-\nu-2(l-1)\mu_B}.
\end{align}
(\ref{lemma6main6}) thus follows from (\ref{lemma63}) and (\ref{lemma64}). Since
\begin{align}\label{lemma6proof10} 
	\|\tilde{\mathbf{D}}^{\rm T}_{[i]}\tilde{\mathbf{D}}_{[i]}\|_F\geq & \sqrt{d	\|\tilde{\mathbf{D}}_{i}\|_2^2} \nonumber \\
	\geq & \sqrt{\frac{d(1-\nu-2(l-1)\mu_B-4\mu_B^2)}{1-\nu-2(l-1)\mu_B}},
\end{align}
and $\|\tilde{\mathbf{D}}_{i}\|_2\leq1$, we arrive (\ref{lemma6main1}). This completes the proof.
\end{IEEEproof}

\subsection{Proof of \textbf{Lemma~\ref{lemma4}}}\label{proofoflemma4} 

\begin{IEEEproof}
Note that $\mathbf{\Omega}_B$ denotes the block index set of $\mathbf{\Omega}$. We need to prove that (\ref{lemma4main1}) and (\ref{lemma4main2}) implies: for the $(l+1)$-th $(0\leq l<k)$ iteration, $\forall j\notin\mathbf{\Omega}_B^{\star}$,
\begin{equation}\label{lemma41} 
	\max_{i\in\mathbf{\Omega}_B^{\star}\backslash\mathbf{\Omega}_B^l}\big\|\tilde{\mathbf{C}}^T_{[i]}\mathbf{R}^l\big\|_{F} > \big\|\tilde{\mathbf{C}}^T_{[j]}\mathbf{R}^l\big\|_{F}.
\end{equation}
	
Since $\mathbf{R}^l=\mathbf{P}^\bot_{\mathbf{D}_{\mathbf{\Omega}^l}}\mathbf{Y}$, we have
\begin{equation}\label{lemma43} 
	\mathbf{R}^l=\mathbf{Q}_{\mathbf{\Omega}_B^l}+\mathbf{P}^\bot_{\mathbf{D}_{\mathbf{\Omega}^l}}\mathbf{N},
\end{equation}
where $\mathbf{Q}_{\mathbf{\Omega}_B^l}=\sum\limits_{i\notin\mathbf{\Omega}_B^l}\tilde{\mathbf{D}}_{[i]}\mathbf{X}_{<i>}$. Following the proofs in \cite{3, 61}, a sufficient condition for (\ref{lemma41}) to hold is: $\forall j\notin \mathbf{\Omega}_B^{\star}$,
\begin{equation}\label{lemma44} 
	\max_{i\in\mathbf{\Omega}_B^{\star}\backslash\mathbf{\Omega}_B^l}\|\tilde{\mathbf{C}}^T_{[i]}\mathbf{Q}_{\mathbf{\Omega}_B^l}\|_{F}-\|\tilde{\mathbf{C}}^T_{[j]}\mathbf{Q}_{\mathbf{\Omega}^l_B}\|_{F}>2\sqrt{d E}\epsilon.
\end{equation}
Let $t\in\arg\max\limits_{i\in\mathbf{\Omega}_B^{\star}\backslash\mathbf{\Omega}^l_B}\|\mathbf{X}_{<i>}\|_F$. Then we have
\begin{align} 
	& \max_{i\in\mathbf{\Omega}_B^{\star}\backslash\mathbf{\Omega}^l_B}\big\|\tilde{\mathbf{C}}^{\rm T}_{[i]}\mathbf{Q}_{\mathbf{\Omega}_B^l}\big\|_{F} \geq \big\|\tilde{\mathbf{C}}^{\rm T}_{[t]}\mathbf{Q}_{\mathbf{\Omega}^l_B}\big\|_{F} \nonumber \\
	& \geq \big\|\tilde{\mathbf{C}}^{\rm T}_{[t]}\tilde{\mathbf{D}}_{[t]}\big|_{F} \|\mathbf{X}_{<t>}\|_F -\!\! \sum_{i\notin\mathbf{\Omega}_B^l\cup\{t\}}\!\! \big\|\tilde{\mathbf{C}}^{\rm T}_{[t]}\tilde{\mathbf{D}}_{[i]}\big\|_F \|\mathbf{X}_{<i>}\|_F \nonumber \\
	& \geq \alpha_l\|\mathbf{X}_{<t>}\|_F  
	  - \beta_l\bigg(\! \sum_{i\in\mathbf{\Omega_B^{\star}}\backslash(\mathbf{\Omega}_B^l\cup\{t\})}\!\!\! \|\mathbf{X}_{<i>}\|_F\! +\!\! \sum_{i\in\bar{\mathbf{\Omega}}_B^{\star}}\!\! \|\mathbf{X}_{<i>}\|_F\bigg), \label{alpha1last}
\end{align}
where (\ref{alpha1last}) follows from \textbf{Lemma~\ref{lemma6}}, and $\forall j\notin\mathbf{\Omega}_B^{\star}$,
\begin{align}\label{lemma6prof2} 
	 \big\|\tilde{\mathbf{C}}^{\rm T}_{[j]}\mathbf{Q}_{\mathbf{\Omega}_B^l}\big\|_{F} & \leq \big\|\tilde{\mathbf{C}}^{\rm T}_{[j]}\tilde{\mathbf{D}}_{[j]}\big\|_{F} \|\mathbf{X}_{<j>}\|_F  
	 + \sum_{i\notin\mathbf{\Omega}^l_B\cup\{j\}} \big\|\tilde{\mathbf{C}}^{\rm T}_{[j]}\tilde{\mathbf{D}}_{[i]}\big\|_F \|\mathbf{X}_{<i>}\|_F \nonumber \\
	 & \leq \|\mathbf{X}_{<j>}\|_F  
	 + \beta_l\bigg(\! \sum_{i\in\mathbf{\Omega_B^{\star}}\backslash\mathbf{\Omega}^l_B}\!\! \|\mathbf{X}_{<i>}\|_F +\!\! \sum_{i\in\bar{\mathbf{\Omega}}_B^{\star}\backslash\{j\}}\!\! \|\mathbf{X}_{<i>}\|_F\bigg) .
\end{align}
Combining (\ref{lemma44}), (\ref{alpha1last}) and (\ref{lemma6prof2}), we have
\begin{align}\label{lemma4last} 
   (\alpha_l+\beta_l)\|\mathbf{X}_{<t>}\|_F - 2\beta_l \sum_{i\in\mathbf{\Omega}_B^{\star}\backslash\mathbf{\Omega}^l_B}\|\mathbf{X}_{<i>}\|_F  
	 > 2\sqrt{d E}\epsilon+(1-\beta_l)\|\mathbf{X}_{<j>}\|_F+2\beta_l\!\! \sum_{i\in\bar{\mathbf{\Omega}}_B^{\star}}\!\! \|\mathbf{X}_{<i>}\|_F.
\end{align}
Since $\|\mathbf{X}_{<j>}\|_F\leq\!\! \sum\limits_{i\in\bar{\mathbf{\Omega}}_B^{\star}}\!\! \|\mathbf{X}_{<i>}\|_F$, the proof is completed. 
\end{IEEEproof}

\subsection{Proof of \textbf{Theorem~\ref{theoremmain1}}}\label{proofoftheorem1} 
	
\begin{IEEEproof}
Since $t\leq l+1$, we have
\begin{equation}\label{theoprof2} 
	\sum_{i\in\mathbf{\Omega}_B^{\star}\backslash\mathbf{\Omega}^l_B}\!\!\! \|\mathbf{X}_{<i>}\|_F\leq \|\mathbf{X}_{<t>}\|_F+(k-l-1)\|\mathbf{X}_{<t+1>}\|_F.
\end{equation}
Therefore, a sufficient condition for (\ref{lemma4main2}) to hold is 
\begin{align}\label{theoprof3} 
	 (\alpha_l-\beta_l)\|\mathbf{X}_{<t>}\|_F - 2\beta_l(k-l-1)\|\mathbf{X}_{<t+1>}\|_F  
	 > 2\bigg(\sqrt{d E}\epsilon + \sum_{i\in\bar{\mathbf{\Omega}}_B^{\star}}\|\mathbf{X}_{<i>}\|_F\bigg). 
\end{align}
On the other hand, it can be derived that $\frac{\beta_l}{\alpha_l-\beta_l}=\frac{1}{\rho-1}$, where $\rho$ is given in (\ref{theomain3}). Hence, (\ref{theoprof3}) can be rewritten as (\ref{theomain2}), and this completes the proof.
\end{IEEEproof}

\subsection{Proof of \textbf{Theorem~\ref{theorem2}}}\label{proofoftheorem2} 

\begin{IEEEproof}
Following the proof of \textbf{Theorem~\ref{theoremmain1}}, we first give the following asymptotic definition:
\begin{equation}\label{theo2prof1} 
  \lim\limits_{\mu_B,\nu\rightarrow0}\rho=\frac{1-\nu-2\theta\mu_B}{\sqrt{d}\nu},
\end{equation}
where $\theta=l-1$. The definition holds due to the high power terms of $\mu_B$ and $\nu$ approaching 0 faster, i.e., $\lim\limits_{\mu_B,\nu\rightarrow0}1-\nu-2(l-1)\mu_B-4\mu_B^2=1-\nu-2(l-1)\mu_B$ and $\lim\limits_{\mu_B,\nu\rightarrow0}\nu-\nu^2-2(l-1)\mu_B\nu+4\mu_B^2=\nu$. Denoting the function $\frac{2(k-\theta)}{\rho-1}$ as $f(\theta)$, we have 
\begin{equation}\label{theo2prof2} 
	\frac{\partial f(\theta)}{\partial\theta} < 0\Rightarrow \nu+\sqrt{d}\nu+2\mu_Bk < 1,
\end{equation}	
which holds with $\mu_B,\nu\rightarrow0$. This means that the condition (\ref{theomain2}) in \textbf{Theorem~\ref{theoremmain1}} can be reformulated into: $\forall j\in\{1,2,\cdots,k\}$,
\begin{align}\label{theo2mainprof1} 
	\|\mathbf{X}_{<t>}\|_F > \frac{2(k-j)\sqrt{d}\nu}{1-\nu-2 j\mu_B-\sqrt{d}\nu}\|\mathbf{X}_{<t+1>}\|_F  
	 + \frac{2\sqrt{d E}\epsilon+(1+\beta_l)\sum_{i\in\bar{\mathbf{\Omega}}_B^{\star}}\|\mathbf{X}_{<i>}\|_F}{\alpha_l-\beta_l}.
\end{align}
Similar to the asymptotic definition in (\ref{theo2prof1}), we have
\begin{align} 
	\lim\limits_{\mu_B,\nu\rightarrow0}\alpha_l-\beta_l =& \sqrt{d}-d\nu,	\label{theo2prof3} \\
	\lim\limits_{\mu_B,\nu\rightarrow0}1+\beta_l =& 1+d\nu. \label{theo2prof33}
\end{align}
The proof is completed by combining (\ref{theo2mainprof1}), (\ref{theo2prof3}) and (\ref{theo2prof33}).
\end{IEEEproof}

\subsection{Proof of \textbf{Lemma~\ref{lemma8}}}\label{proofoflemma8} 

\begin{IEEEproof}
Note that the exact recovery condition in \cite{4} applies to the S-BOMP/S-BOLS/FS-BOLS algorithms, which indicates that: if (\ref{reconstructible1}) holds, the algorithms perform exact recovery in the noiseless case. Then, following the proof of \textbf{Lemma~\ref{lemma4}}, we obtain the inequality
\begin{align}\label{coro3main1} 
 \max_{i\in\mathbf{\Omega}_B^{\star}\backslash\mathbf{\Omega}_B^l}\big\|\tilde{\mathbf{C}}^{\rm T}_{[i]}\mathbf{Q}_{\mathbf{\Omega}_B^l}\big\|_{F}  
	 > 2\frac{1-(d-1)\nu-(k-1)d\mu_B}{1-(d-1)\nu-(2k-1)d\mu_B}\|\mathbf{N}_{i,\mathbf{\Omega}^l}\|_F,
\end{align}
which ensures that the algorithms select a correct support in the current iteration, and the inequality (\ref{coro3main1}) is derived based on \cite{3,4}. On the other hand, 
\begin{align}\label{coro3main2} 
 	 \max_{i\in\mathbf{\Omega}_B^{\star}\backslash\mathbf{\Omega}_B^l}\big\|\tilde{\mathbf{C}}^{\rm T}_{[i]}\mathbf{Q}_{\mathbf{\Omega}_B^l}\big\|_{F}  
 	 = \max_{i\in\mathbf{\Omega}^{\star}\backslash\mathbf{\Omega}}\sqrt{\sum_{j\in\mathbf{\Omega}_B^{\star}\backslash\mathbf{\Omega}_B^l}\big\|\tilde{\mathbf{C}}^{\rm T}_{[i]}\mathbf{P}^{\bot}_{\mathbf{D}_{\mathbf{\Omega}^l}}\mathbf{D}_{\mathbf{\Omega}^{\star}\backslash\mathbf{\Omega}^l}\mathbf{X}_{<j>}\big\|^2_{F}}  
 	 \geq \alpha_l\sqrt{\sum_{j\in\mathbf{\Omega}_B^{\star}\backslash\mathbf{\Omega}_B^l}\|\mathbf{X}_{<j>}\|^2_{F}}.
\end{align}
This and the inequality (\ref{coro3main1}) indicate that a sufficient condition for selecting a correct support in the $(l+1)$-th iteration is (\ref{lemma8main1}).
\end{IEEEproof}

\subsection{Proof of \textbf{Theorem~\ref{theo3}}}\label{proofoftheorem3} 

\begin{IEEEproof}
Based on the assumption $\forall e\in\{1,2,\cdots,E\}$, $\|\mathbf{N}_e\|_2\leq\epsilon$, the following inequality holds:
\begin{equation}\label{theo3prof1} 
	\big\|\mathbf{P}^{\bot}_{\mathbf{D}_{\mathbf{\Omega}^l}}\mathbf{N}\big\|_F = \sqrt{\sum_{e=1}^{E}\big\|\mathbf{P}^{\bot}_{\mathbf{D}_{\mathbf{\Omega}^l}}\mathbf{N}_e\big\|^2_2}\leq\sqrt{E}\epsilon.
\end{equation}
On the other hand, 
\begin{equation}\label{theo3prof2} 
	\|\mathbf{N}_{i,\mathbf{\Omega}}\|_F\leq \big\|\tilde{\mathbf{C}}^{\rm T}_{[i]}\big\|_F	\big\|\mathbf{P}^{\bot}_{\mathbf{D}_{\mathbf{\Omega}^l}}\mathbf{N}\big\|_F\leq \sqrt{d E}\epsilon,
\end{equation}
where $\|\mathbf{N}_{i,\mathbf{\Omega}}\|_F$ is defined in \textbf{Lemma~\ref{lemma8}}. Therefore, 
\begin{align}\label{theo3prof3} 
	 \sqrt{\sum_{j\in\mathbf{\Omega}_B^{\star}\backslash\mathbf{\Omega}_B^l}\|\mathbf{X}_{<j>}\|^2_{F}}  
	  > 
	  2\frac{1-(d-1)\nu-(k-1)d\mu_B}{\alpha_l(1-(d-1)\nu-(2k-1)d\mu_B)}\sqrt{d E}\epsilon
\end{align}
is a sufficient condition for the algorithms to select a correct block. As the data length $N\rightarrow\infty$, according to the central limit theorem, the atoms in $\mathbf{X}_{<j>}$, $j\in\mathbf{\Omega}_B^{\star}\backslash\mathbf{\Omega}^l_B$, are i.i.d. with the Gaussian distribution $\mathcal{N}(0,1)$. Then $\sum_{j\in\mathbf{\Omega}_B^{\star}\backslash\mathbf{\Omega}_B^l}\|\mathbf{X}_{<j>}\|^2_{F}$ is a $\chi^2_{(k-l)d E}$ random variable. Based on \cite[Lemma 3]{11}, we have
\begin{equation}\label{theo3prof4} 
	\Pr\Bigg(	\sqrt{\sum_{j\in\mathbf{\Omega}_B^{\star}\backslash\mathbf{\Omega}_B^l}\|\mathbf{X}_{<j>}\|^2_{F}}\geq \gamma\Bigg) \geq 1-\frac{1}{(k-l)d E} .
\end{equation}
By combining (\ref{theo3prof3}) and (\ref{theo3prof4}), the proof is completed.
\end{IEEEproof}
	
\subsection{Proof of \textbf{Theorem~\ref{theo4}}}\label{proofoftheorem4} 

\begin{IEEEproof}
Observe that 
\begin{equation}\label{theo4prof1} 
  \sqrt{\sum\nolimits_{j\in\mathbf{\Omega}_B^{\star}\backslash\mathbf{\Omega}_B^l}\|\mathbf{X}_{<j>}\|^2_{F}}\geq\sqrt{E(k-l)}s.
\end{equation}
On the other hand,
\begin{align}\label{theo4prof2} 
 	\|\mathbf{N}_{i,\mathbf{\Omega}^l}\|_F =	\big\|\tilde{\mathbf{C}}^{\rm T}_{[i]}\mathbf{P}^{\bot}_{\mathbf{D}_{\mathbf{\Omega}^l}}\mathbf{N}\big|_F \leq \big\|\tilde{\mathbf{C}}^{\rm T}_{[i]}\big|_F\big\|\mathbf{P}^{\bot}_{\mathbf{D}_{\mathbf{\Omega}^l}}\mathbf{N}\big\|_F  
 	\leq  \sqrt{d}\big\|\mathbf{P}^{\bot}_{\mathbf{D}_{\mathbf{\Omega}^l}}\mathbf{N}\big\|_F.
\end{align}
Note that the variable $\|\mathbf{P}^{\bot}_{\mathbf{D}_{\mathbf{\Omega}^l}}\mathbf{N}\|_F^2/\sigma^2$ follows the chi-squared distribution with $(k-l)Ed$ degrees of freedom. Based on \cite[Lemma 5.1]{10}, we have
\begin{equation}\label{theo4prof3} 
	\Pr\Big(\big\|\mathbf{P}^{\bot}_{\mathbf{D}_{\mathbf{\Omega}^l}}\mathbf{N}\big\|_F\leq\zeta_l\Big)\geq 1-\frac{1}{(k-l)d E}.
\end{equation}
Therefore,
\begin{equation}\label{theo4prof4} 
	\sqrt{E(k-l)}s\geq \sqrt{d}\zeta_l
\end{equation}
is a sufficient condition that ensures the algorithms to select a correct support in the $(l+1)$-th iteration with the probability at least $1-\frac{1}{(k-l)d E}$. This completes the proof.
 \end{IEEEproof}

\subsection{Proof of \textbf{Corollary~\ref{coro2}}}\label{proofofcorollary1} 

\begin{IEEEproof}
Since 
\begin{align}\label{coro2prof1} 
	\|\mathbf{N}_{i,\mathbf{\Omega}^l}\|_F = \big\|\tilde{\mathbf{C}}^{\rm T}_{[i]}\mathbf{P}^{\bot}_{\mathbf{D}_{\mathbf{\Omega}^l}}\mathbf{N}\big\|_F \geq \sqrt{\lambda_{\min}\big(\mathbf{D}^{\rm T}_{[i]}\mathbf{P}^{\bot}_{\mathbf{D}_{\mathbf{\Omega}^l}}\mathbf{D}_{[i]}\big)}\|\mathbf{N}\|_F  
	\geq  \sqrt{1-(d-1)\nu}\|\mathbf{N}\|_F,
\end{align}
where the last inequality is based on \textbf{Lemma~\ref{lemma1}}, the condition in (\ref{lemma8main1}) changes into
\begin{align}\label{coro2prof2} 
	\sqrt{\sum_{j\in\mathbf{\Omega}_B^{\star}\backslash\mathbf{\Omega}_B^l}\|\mathbf{X}_{<j>}\|^2_{F}} > 2\frac{1-(d-1)\nu-(k-1)d\mu_B}{\alpha_l(1-(d-1)\nu-(2k-1)d\mu_B)}  
	 \times \sqrt{1-(d-1)\nu}\|\mathbf{N}\|_F.
\end{align}
It can be observed that the two chi-squared random variables $\sum_{j\in\mathbf{\Omega}_B^{\star}\backslash\mathbf{\Omega}_B^l}\|\mathbf{X}_{<j>}\|^2_{F}$ and $\|\mathbf{N}\|_F$ are independent. 

Hence, similar to the proofs of \textbf{Theorems~\ref{theo3}} and \textbf{\ref{theo4}}, we conclude that (\ref{coro2main}) is a sufficient condition for the algorithms to select a correct support in the $(l+1)$-th iteration with the probability at least $1-\frac{1}{(k-l)d E}-\frac{1}{k d E}$.
\end{IEEEproof}

\subsection{Proof of \textbf{Theorem~\ref{theo5}}}\label{proofoftheorem5} 

\begin{IEEEproof}
Based on \textbf{Corollary~\ref{coro2}}, (\ref{analysis222}) and (\ref{theo3prof6}), we conclude that
\begin{align}\label{theo555prof1} 
	 \xi\big(\sqrt{4d E-2} - \sqrt{d E+2\sqrt{d E\log(d E)}}\big)  
	  > \frac{1}{\sqrt{d}}\mathcal{T}\sigma \sqrt{k Ed +2\sqrt{k E d\log(k E d)}}
\end{align}
is a sufficient condition for the S-BOMP/S-BOLS/FS-BOLS algorithms to select $k$ correct supports during $k$ iterations. Since $k\leq d$ and the inequality $d^2 E\log(d^2 E)\leq d^3 E\log(d E)$ holds, a sufficient condition for (\ref{theo555prof1}) to hold is
\begin{align}\label{theo555prof3} 
	 \xi\big(\sqrt{4d^2E-2d}-\sqrt{d^2E+2\sqrt{d^3E\log(d E)}}\big)  
	  > \mathcal{T}\sigma\sqrt{d^2E+2\sqrt{d^3E\log(d E)}}.
\end{align}
Noticing that $d E\geq 1$ and $\log (d)\leq d E-1$, the following inequality is sufficient for the establishment of (\ref{theo555prof3}):
\begin{equation}\label{theo555prof4} 
	a_1E^2+b_1E+c_1>0,
\end{equation}
where $a_1$, $b_1$ and $c_1$ are given in (\ref{eq-a1}) to (\ref{eq-c1}), and (\ref{theo5main}) is the portion of the resolution of (\ref{theo555prof4}) that is not less than 1. 
\end{IEEEproof}

\ifCLASSOPTIONcaptionsoff
  \newpage
\fi


\end{document}